\def\psra{PSR J0737--3039A}
\def\psrb{PSR J0737--3039B}
\def\psrab{PSR J0737--3039A/B}

\def\nk{n_{\rm b}}

\def\Pb{P_{\rm b}}

\def\rfr#1{Equation\,(\ref{#1})}
\def\rfrs#1#2{Equations\,(\ref{#1})-(\ref{#2})}
\def\Rfr#1{Equation\,(\ref{#1})}
\def\Rfrs#1#2{Equations\,(\ref{#1})-(\ref{#2})}

\def\virg#1{``#1"}

\def\eqi{\begin{equation}}
\def\eqf{\end{equation}}
\def\eqia{\begin{eqnarray}}
\def\eqfa{\end{eqnarray}}

\def\rp#1#2{{#1\over#2}}
\def\lb#1{\label{#1}}

\def\bds#1{\boldsymbol{#1}}

\def\ton#1{\left(#1\right)}
\def\qua#1{\left[#1\right]}
\def\grf#1{\left\{#1\right\}}

\documentclass[linenumbers]{aastex}

\usepackage{morefloats}
\usepackage[title]{appendix}
\usepackage{hyperref,textcomp}
\usepackage{booktabs}
\usepackage[table,xcdraw]{xcolor}
\usepackage{multirow}
\usepackage{rotating,tabularx}
\usepackage{float}
\usepackage{enumerate}
\usepackage{rotating}
\usepackage[polutonikogreek,english]{babel}
\usepackage{amsmath,starfont,textgreek,w-greek,wasysym}
\usepackage[flushleft]{threeparttable}
\usepackage{amsthm}
\usepackage{amscd,lineno}
\usepackage{amssymb,dsfont}
\usepackage{graphicx,epsfig}
\usepackage[utf8]{inputenc}
\usepackage[T1]{fontenc}
\usepackage{txfonts}
\usepackage{xr-hyper}

\RequirePackage{color}

\newcommand{\emaila}{lorenzo.iorio@libero.it}

\allowdisplaybreaks[1]

\begin{document}

\title{The impact of the spin--orbit misalignment and of the spin of B  on the Lense--Thrirring orbital precessions of the Double
Pulsar PSR J0737--3039A/B}

\shortauthors{L. Iorio}

\author{Lorenzo Iorio\altaffilmark{1} }
\affil{Ministero dell'Istruzione, dell'Universit\`{a} e della Ricerca
(M.I.U.R.)
\\ Viale Unit\`{a} di Italia 68, I-70125, Bari (BA),
Italy}

\email{\emaila}

\begin{abstract}
In the Double Pulsar, the Lense--Thirring periastron precession $\dot\omega^\mathrm{LT}$  could  be used to measure/constrain the moment of inertia $\mathcal{I}_\mathrm{A}$ of A. Conversely, if $\mathcal{I}_\mathrm{A}$ will be independently determined with sufficient accuracy by other means, tests of the Lense--Thirring effect could be performed. Such findings rely upon a formula for $\dot\omega^\mathrm{LT,\,A}$ induced by the spin angular momentum ${\bds S}^\mathrm{A}$ of A, valid if the orbital angular momentum $\bds L$ and ${\bds S}^\mathrm{A}$ are aligned, and neglecting  $\dot\omega^\mathrm{LT,\,B}$ because of the smallness of  ${\bds S}^\mathrm{B}$. The impact on $\dot\omega^\mathrm{LT,\,A}$ of the departures of the ${\bds S}^\mathrm{A}$-$\bds L$ geometry from the ideal alignment is calculated. With the current upper bound on the misalignment angle $\delta_\mathrm{A}$ between them, the angles $\lambda_\mathrm{A},\,\eta_\mathrm{A}$ of ${\bds S}^\mathrm{A}$ are constrained within $85^\circ \lb{lam} \lesssim \lambda_\mathrm{A}\lesssim 92^\circ,\, 266^\circ \lesssim \eta_\mathrm{A} \lesssim 274^\circ$. In units of the first order post-Newtonian mass-dependent periastron precession $\dot\omega^\mathrm{GR}=16.89^\circ\,\mathrm{yr}^{-1}$, a range variation $\Delta\dot\omega^\mathrm{LT,\,A}\doteq\dot\omega^\mathrm{LT,\,A}_\mathrm{max} - \dot\omega^\mathrm{LT,\,A}_\mathrm{min} = 8\times 10^{-8}\,\omega^\mathrm{GR}$ is implied. The experimental uncertainty $\sigma_{\dot\omega_\mathrm{obs}}$ in measuring the periastron rate should become smaller by 2028-2030. Then, the spatial orientation of ${\bds{S}}^\mathrm{B}$ is constrained from the existing bounds on the misalignment angle $\delta_\mathrm{B}$, and  $\dot\omega^\mathrm{LT,\,B}\simeq 2\times 10^{-7}\,\dot\omega^\mathrm{GR}$ is correspondingly calculated. The error $\sigma_{\dot\omega_\mathrm{obs}}$ should become smaller around 2025. The Lense--Thirring inclination and node precessions $\dot I^\mathrm{LT},\,\dot \Omega^\mathrm{LT}$ are predicted to be $\lesssim 0.05\,\mathrm{arcseconds\,per\,year}$, far below the current experimental accuracies $\sigma_{I_\mathrm{obs}}=0.5^\circ,\,\sigma_{\Omega_\mathrm{obs}}=2^\circ$ in measuring $I,\,\Omega$ over $1.5$ year with the scintillation technique. The Lense--Thirring rate $\dot x_\mathrm{A}^\mathrm{LT}$ of the projected semimajor axis $x_\mathrm{A}$ of \psra\, is $\lesssim 2\times 10^{-16}\,\mathrm{s\,s}^{-1}$, just two orders of magnitude smaller than a putative experimental uncertainty $\sigma_{\dot x^\mathrm{obs}_\mathrm{A}}\simeq 10^{-14}\,\mathrm{s\,s}^{-1}$ guessed from 2006 data.
\end{abstract}


%
%


\section{Introduction}
The double pulsar \psrab, discovered in 2003  \citep{2003Natur.426..531B,2004Sci...303.1153L}, is a tight binary system  made of two neutron stars, \psra\,and \psrb, completing a mildly eccentric orbit in $2.45\,\mathrm{hr}$. A distinctive feature of such a system is that, at least for some years (2003-2008), both its components were simultaneously detectable as emitting radio pulsars. It is currently possible to collect pulses only from \psra\, because  the general relativistic geodetic precession \citep{1974CRASM.279..971D,1975PhRvD..12..329B} of the spin of \psrb, measured in 2008 by \citet{2008Sci...321..104B} to a $\simeq 13\%$ accuracy, displaced its radio beam away from the line of sight. Since its discovery, \psrab\, turned out to be a unique laboratory to perform tests of relativistic gravity in a stronger regime than in our solar system \citep{2006Sci...314...97K}.


To the first order of its post-Newtonian (1pN) expansion, the General Theory of Relativity (GTR) predicts, among other things, that the argument of periastron $\omega$ of the relative orbit of a gravitationally bound binary system made of two spinning bodies A and B with masses $M_\mathrm{A},\,M_\mathrm{B}$ and angular momenta ${\bds S}^\mathrm{A},\,{\bds S}^\mathrm{B}$ undergoes an orbit--averaged, long term variation  made of two contributions. The first one, dubbed as \virg{gravitoelectric} \citep{2001rfg..conf..121M,Mash03} and denoted here as $\dot\omega^\mathrm{GR}$, depends only on the sum $M$ of the masses of A and B. It is the generalization of the time-honored, formerly anomalous perihelion precession of Mercury in the field of the Sun explained by \citet{Ein15} with his GTR, and reads \citep{Rob38,1988NCimB.101..127D,Sof89,1991ercm.book.....B}
\eqi
\dot\omega^\mathrm{GR} = \rp{3\,\nk\,\mu}{c^2\,a\,(1-e^2)}\lb{omGR}.
\eqf
In \rfr{omGR}, $c$ is the speed of light in vacuum, $\mu\doteq GM$ is the gravitational parameter of the binary  made of the product of the Newtonian constant of gravitation $G$ times  $M$, $a$ and $e$ are the semimajor axis and the eccentricity, respectively, of the relative orbit, while $\nk\doteq\sqrt{\mu/a^3}$ is the Keplerian mean motion.

The second general relativistic contribution to the 1pN periastron precession, dubbed as \virg{gravitomagnetic} \citep{Thorne86,1986hmac.book..103T,1988nznf.conf..573T,2001rfg..conf..121M,2001rsgc.book.....R,Mash03} and denoted here as $\dot\omega^\mathrm{LT}$,  depends on both the masses and the spins of A and B, and is the generalization to the two-body case of the\footnote{According to the historical analyses by \citet{2007GReGr..39.1735P,2008mgm..conf.2456P,Pfister2014}, it should be dubbed, more appropriately, as Einstein--Thirring--Lense effect.} Lense--Thirring effect, originally worked out by \citet{LT18,1984GReGr..16..727M} for a point particle orbiting a massive rotating primary, the orientation in space of whose spin is assumed to be known. For an arbitrary orientation in space of the spins of the binary's components, also the inclination $I$ and the node $\Omega$ experience long--term gravitomagnetic spin-orbit shifts. Contrary to what sometimes misrepresented in the literature \citep{2020MNRAS.497.3118H}, the only unquestioned test of gravitomagnetism  performed, to date, in the weaker field of Earth was carried out with the dedicated space-based mission Gravity Probe B (GP-B) \citep{Varenna74} measuring the Pugh-Schiff spin precessions \citep{Pugh59,Schiff60} of four spaceborne gyroscopes to a $19$ percent accuracy \citep{2011PhRvL.106v1101E,2015CQGra..32v4001E} instead of the originally envisaged $\simeq 1$ percent \citep{2001LNP...562...52E}. The attempts to measure the Lense--Thirring orbital precessions with the Earth's artificial satellites of the LAGEOS type \citep{2013NuPhS.243..180C,2019Univ....5..141L,2020Univ....6..139L} have always been controversial so far; see, e.g., \citet{2011Ap&SS.331..351I,2013CEJPh..11..531R,2013NewA...23...63R,2014NewA...29...25R}, and references therein.

In principle, also the quadrupole mass moments $Q^\mathrm{A},\,Q^\mathrm{B}$ of A and B has an impact on the shifts of $I,\,\Omega$ and $\omega$ through formally Newtonian contributions\footnote{In fact, the are of the order of $\mathcal{O}\ton{c^{-2}}$ because of the relativistic expressions of the quadrupole mass moments of highly compact objects such as neutron stars.}  which, in the present case, will turn out to be negligible \citep{2020MNRAS.497.3118H}, as it will be independently confirmed here.

To the 2pN order, there is another gravitoelectric contribution to the periastron precession depending only on the masses of A and B \citep{1988NCimB.101..127D,1993PhLA..174..196S,1993PhLA..177..461S}; it will not be treated here.

The extraction of $\dot\omega^\mathrm{LT}$ from the experimentally measured total periastron precession $\dot\omega_\mathrm{obs}$ of \psrab, a possibility firstly envisaged by \citet{2004Sci...303.1153L,2005ApJ...629..979L,2006Sci...314...97K}, would allow to get important insights on the equation of state (EOS) of the dense matter inside neutron stars. Indeed, by assuming the validity of GTR, the knowledge of $\dot\omega^\mathrm{LT}$ could be used to constrain the EOS through the determination of the moment of inertia (MOI) $\mathcal{I}_\mathrm{A}$ of \psra. Conversely, if the MOI could be independently determined by other means, a test of the Lense--Thirring effect could be performed to some level of accuracy. In this respect, a necessary condition for the successful outcome of the aforementioned strategy is that the experimental error $\sigma_{\dot\omega_\mathrm{obs}}$ in determining the periastron precession is adequately smaller than $\dot\omega^\mathrm{LT}$. Furthermore,  as quantitatively investigated  for the first time by \citet{2009NewA...14...40I}, \rfr{omGR} must be calculated with sufficient accuracy in order to be subtracted from $\dot\omega_\mathrm{obs}$, acting as a source of major systematic uncertainty; see also the general discussion by \citet{1988NCimB.101..127D} before the discovery of \psrab.
The same holds also for the 2pN contribution to the periastron precession whose magnitude may be comparable to the 1pN gravitomagnetic one \citep{2020MNRAS.497.3118H}.
To this aim, the masses of both the neutron stars must be independently and accurately determined. Recently, \citet{2020MNRAS.497.3118H} thoroughly investigated such aspects in view of the increasing amount of accurate pulsar timing data which will be collected from the MeerKAT and SKA facilities in the ongoing decade. For a previous analysis, see \citet{Kehletal017}; \citet{2021arXiv210705812M} dealt with binary pulsars with shorter orbital periods, yet to be discovered. Basing their analyses on the contribution to $\dot\omega^\mathrm{LT}$ from \psra\, only, \citet{2020MNRAS.497.3118H} concluded that a MOI measurement with 11 per cent accuracy (68 per cent confidence) would be possible by 2030. Conversely, by assuming a sufficiently accurate knowledge of the EOS by that date, \citet{2020MNRAS.497.3118H} suggested that a Lense--Thirring test accurate to the 7 per cent level could be feasible. \citet{2020MNRAS.497.3118H} neglected the contribution of  ${\bds S}^\mathrm{B}$  to $\dot\omega^\mathrm{LT}$ because of the much slower rotation of \psrb\, with respect to \psra. Moreover, in considering only $\dot\omega^\mathrm{LT,\,A}$, \citet{2020MNRAS.497.3118H} assumed ${\bds S}^\mathrm{A}$ exactly aligned with the orbital angular momentum $\bds L$.

The paper is organized as follows.
In Section\,\ref{sec1},  such assumptions by \citet{2020MNRAS.497.3118H} are quantitatively checked in order to see if the full expression of $\dot\omega^\mathrm{LT}$, compared with the expected future improvements in $\sigma_{\dot\omega_\mathrm{obs}}$, will be, sooner or later, required in the timing of \psrab. In particular, the consequences on $\dot\omega^\mathrm{LT,\,A}$ of the misalignment between $\bds L$ and ${\bds S}^\mathrm{A}$ are treated in Section\,\ref{sec1.1}, while the contribution of ${\bds S}^\mathrm{B}$ to $\dot\omega^\mathrm{LT}$ is evaluated in Section\,\ref{sec1.2}. The gravitomagnetic spin-orbit precessions of $I$ and $\Omega$ are dealt with in Section\,\ref{sec2}: their upper bounds are calculated and compared with the current experimental uncertainties in determining such orbital elements.
The long-term rates of change of $I,\,\Omega$ and $\omega$ due to the quadrupole mass moments of A and B are worked out in Section\,\ref{sec3}.
Section\,\ref{sec4} summarizes the findings of the paper and offers concluding remarks.
\section*{Notations}
Here, some basic notations and definitions used in the text are shown
\begin{description}
  \item[] $G:$\,Newtonian constant of gravitation
  \item[] $c:$\,speed of light in vacuum
  \item[] $M_\mathrm{A}:$\,mass of \psra
  \item[] $\mathcal{I}_\mathrm{A}:$\, moment of inertia (MOI) of \psra
  \item[] $\nu_\mathrm{A}:$\, spin frequency of \psra
  \item[] $P_\mathrm{A}:$\, spin period of \psra
  \item[] $Q_\mathrm{A}:$\, dimensional quadrupole mass moment of \psra
  \item[] ${\boldsymbol{S}}^\mathrm{A}:$\, spin angular momentum of \psra
  \item[] ${\boldsymbol{\hat{S}}}^\mathrm{A}:$\, unit vector of the spin angular momentum of \psra
  \item[] $S^\mathrm{A}=\mathcal{I}_\mathrm{A}\,2\uppi\nu_\mathrm{A}:$\, magnitude of the spin angular momentum of \psra
  \item[] $\lambda_\mathrm{A}:$ angle between the reference $z$ axis, directed along the line of sight away from the observer, and the spin axis of \psra\,\citep[][Figure\,1]{1992PhRvD..45.1840D}
  \item[] $\psi_\mathrm{A}:$ angle between the reference $x$ axis and the projection of the spin axis of \psra\,onto the plane of the sky, assumed as reference $\grf{x,\,y}$ plane \citep[][Figure\,1]{1992PhRvD..45.1840D}
 \item[] $M_\mathrm{B}:$\,mass of \psrb
  \item[] $\mathcal{I}_\mathrm{B}:$\, moment of inertia (MOI) of \psrb
  \item[] $\nu_\mathrm{B}:$\, spin frequency of \psrb
  \item[] $P_\mathrm{B}:$\, spin period of \psrb
  \item[] $Q_\mathrm{B}:$\, dimensional quadrupole mass moment of \psrb
  \item[] ${\boldsymbol{S}}^\mathrm{B}:$\, spin angular momentum of \psrb
  \item[] ${\boldsymbol{\hat{S}}}^\mathrm{B}:$\, unit vector of the spin angular momentum of \psrb
  \item[] $S^\mathrm{B}=\mathcal{I}_\mathrm{B}\,2\uppi\nu_\mathrm{B}:$\, magnitude of the spin angular momentum of \psrb
  \item[] $\lambda_\mathrm{B}:$ angle between the reference $z$ axis, directed along the line of sight away from the observer, and the spin axis of \psrb\,\citep[][Figure\,1]{1992PhRvD..45.1840D}
  \item[] $\psi_\mathrm{B}:$ angle between the reference $x$ axis and the projection of the spin axis of \psrb\,onto the plane of the sky, assumed as reference $\grf{x,\,y}$ plane \citep[][Figure\,1]{1992PhRvD..45.1840D}
  \item[] $M\doteq M_\mathrm{A} + M_\mathrm{B}:$\, total mass of \psrab
  \item[] $\mu\doteq GM:$\, gravitational parameter of \psrab
  \item[] $a:$\, semimajor axis of the relative orbit of \psrab
  \item[] $\nk\doteq\sqrt{\mu/a^3}:$\, mean motion
  \item[] $\Pb\doteq 2\uppi/\nk:$\, orbital period
  \item[] $a_\mathrm{A}= a\,M_\mathrm{B}/M:$\, barycentric semimajor axis of \psra
  \item[] $e:$\, orbital eccentricity
  \item[] ${\boldsymbol{L}}:$\, orbital angular momentum
  \item[] $\delta_\mathrm{A}:$\, misalignment angle between $\bds L$ and ${\bds S}^\mathrm{A}$
   \item[] $\delta_\mathrm{B}:$\, misalignment angle between $\bds L$ and ${\bds S}^\mathrm{B}$
  \item[] $I:$\, angle between the reference $z$ axis, directed along the line of sight away from the observer, and the orbital angular momentum \citep[][Figure\,1]{1992PhRvD..45.1840D}
 \item[] $x_\mathrm{A}\doteq a_\mathrm{A}\sin I/c:$\, projected semimajor axis of \psra
  \item[] $\Omega:$\, longitude of the ascending node
  \item[] $\eta_\mathrm{A}\doteq \psi_\mathrm{A}-\Omega:$\, angle between the projection of the spin axis of \psra\,onto the plane of the sky and the longitude of the ascending node \citep[][Figure\,1]{1992PhRvD..45.1840D}
  \item[] $\eta_\mathrm{B}\doteq \psi_\mathrm{B}-\Omega:$\, angle between the projection of the spin axis of \psrb\,onto the plane of the sky and the longitude of the ascending node \citep[][Figure\,1]{1992PhRvD..45.1840D}
  \item[] $\boldsymbol{n}\doteq\grf{\sin I\sin\Omega,\,-\sin I\cos\Omega,\,\cos I}:$\, unit vector of the orbital angular momentum
  \item[] $\boldsymbol{l}\doteq\grf{\cos\Omega,\,\sin\Omega,\,0}:$\, unit vector directed along the line of the nodes towards the ascending node
  \item[] $\boldsymbol{m}\doteq\grf{-\cos I\sin\Omega,\,\cos I\cos\Omega,\,\sin I}:$\, unit vector in the orbital plane perpendicular to $\boldsymbol{l}$ such that $\boldsymbol{l}\boldsymbol\times\boldsymbol{m} = \boldsymbol{n}$.
  \item[] $\omega:$\, argument of periastron
\end{description}
For the numerical values of most of such quantities, see \citet{2006Sci...314...97K}.
\section{Using the periastron precession}\lb{sec1}
In the following, the coordinate system of Figure\,1 in \citet{1992PhRvD..45.1840D} will be adopted; its reference $z$ axis is directed along the line of the sight away from the observer, while the reference $\grf{x,\,y}$ plane is assumed coincident with the plane of the sky. Thus, the components of, say, ${\bds{\hat{S}}}^\mathrm{A}$ can be parameterized in terms of the spherical angles $\lambda_\mathrm{A},\,\psi_\mathrm{A}$ as
\begin{align}
{\hat{S}}_x^\mathrm{A} & = \sin\lambda_\mathrm{A}\cos\psi_\mathrm{A}, \\ \nonumber \\
{\hat{S}}_y^\mathrm{A} & = \sin\lambda_\mathrm{A}\sin\psi_\mathrm{A}, \\ \nonumber \\
{\hat{S}}_z^\mathrm{A} & = \cos\lambda_\mathrm{A}.
\end{align}

The full analytical expression of the Lense--Thirring periastron precession of a two--body system whose spin angular momenta are arbitrarily oriented in space is \citep{2017EPJC...77..439I}
\begin{align}
\dot\omega^\mathrm{LT} &= -\rp{2\,G\,S^\mathrm{A}}{c^2\,a^3\,\ton{1-e^2}^{3/2}}\ton{1+\rp{3}{4}\rp{M_\mathrm{B}}{M_\mathrm{A}}}{\bds{\hat{S}}}^\mathrm{A}\bds\cdot\ton{2\bds n + \cot I\,\bds m } + \mathrm{A}\leftrightarrows\mathrm{B}=\\ \nonumber \\
&= -\rp{G\,S^\mathrm{A}}{c^2\,a^3\,\ton{1-e^2}^{3/2}}\ton{1+\rp{3}{4}\rp{M_\mathrm{B}}{M_\mathrm{A}}}\qua{6\cos I\cos\lambda_\mathrm{A}+\ton{3\cos 2I - 1}\csc I\sin\lambda_\mathrm{A}\sin\eta_\mathrm{A}} + \mathrm{A}\leftrightarrows\mathrm{B}.\lb{omeLT}
\end{align}
For earlier, more or less explicit results, see, e.g., \citet{Kali59}, \citet{Micha60}, \citet[][Equations\,(5.11a)]{1988NCimB.101..127D}, \citet[][Section\,4.4.4]{1991ercm.book.....B} and \citet{1995CQGra..12..983W}. Furthermore, in \citet{1975PhRvD..12..329B}  the two-body Lense--Thirring orbital precessions are written in vectorial form.
If ${\bds{\hat{S}}}^\mathrm{A}$ and $\bds n$ are exactly aligned, i.e. if
\begin{align}
I &= \lambda_\mathrm{A},\\ \nonumber \\
\eta_\mathrm{A} &= 270^\circ,
\end{align}
from \rfr{omeLT} one has
\eqi
\dot\omega^\mathrm{LT,\,A} = -\rp{4\,G\,S^\mathrm{A}}{c^2\,a^3\,\ton{1-e^2}^{3/2}}\ton{1+\rp{3}{4}\rp{M_\mathrm{B}}{M_\mathrm{A}}}.\lb{lineare}
\eqf
\subsection{The impact of the misalignment between ${\bds S}^\mathrm{A}$ and $\bds L$}\lb{sec1.1}
Relying upon existing constraints on $\delta_\mathrm{A}$ pointing towards a small misalignment between the orbital angular momentum and the spin axis of \psra, \citet{2020MNRAS.497.3118H} based their simulations on the assumption of an ideally perfect alignment between $\bds n$ and ${\bds{\hat{S}}}^\mathrm{A}$. Here,  the impact on $\dot\omega^\mathrm{LT,\,A}$ of the departures from such a necessarily idealized condition will be quantitatively assessed.

About the MOI of \psra, estimates for it were recently given in the literature; see \citet{2021PhRvL.126r1101S} and references therein.
Relying upon such results, the value
\eqi
\mathcal{I}_\mathrm{A} = 1.6\times 10^{38}\,\mathrm{kg\,m}^2 \lb{ine}
\eqf
is assumed here.

\begin{figure}[H]
\centering
\centerline{
\vbox{
\begin{tabular}{c}
\epsfxsize= 16 cm\epsfbox{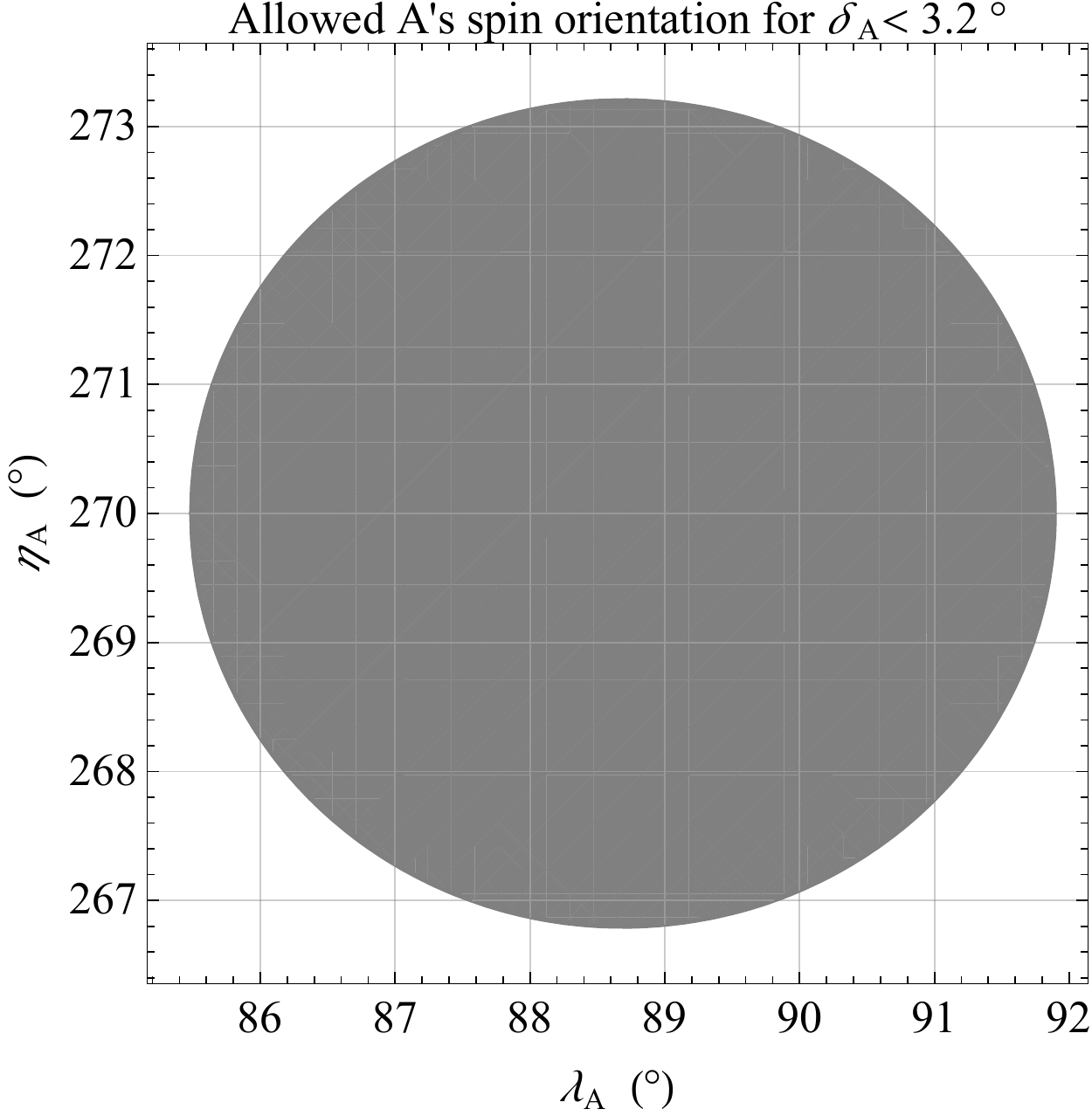}\\
\end{tabular}
}
}
\caption{
Allowed region, in the $\grf{\lambda_\mathrm{A},\,\eta_\mathrm{A}}$ plane, for the position angles $\lambda_\mathrm{A},\,\eta_\mathrm{A}$ of the spin ${\bds S}^\mathrm{A}$ of \psra. It was obtained by plotting $\delta_\mathrm{A}$, calculated with $I=88.69^\circ$ \citep{2006Sci...314...97K}, as a function of  $\lambda_\mathrm{A},\,\eta_\mathrm{A}$ according to \rfr{angolo}, and imposing the constraint of \rfr{uper}.
}\label{fig1}
\end{figure}

The misalignment angle $\delta_\mathrm{A}$  is a function of $I,\,\lambda_\mathrm{A},\,\eta_\mathrm{A}$ defined by
\eqi
\cos \delta_\mathrm{A} = \bds n\bds\cdot{\bds{\hat{S}}}^\mathrm{A}= \cos I\cos\lambda_\mathrm{A} - \sin I\sin\eta_\mathrm{A}\sin\lambda_\mathrm{A}.\lb{angolo}
\eqf
According to \citet{2013ApJ...767...85F}, who analysed the pulse profile shape over six years, it is close to zero, with an upper bound as little as
\eqi
\delta_\mathrm{A}\leq 3.2^\circ.\lb{uper}
\eqf
The bound of \rfr{uper} was inferred by  assuming that the observed emission comes from both magnetic poles \citep{2013ApJ...767...85F}.
By adopting the best estimate
\eqi
I =88.69^\circ
\eqf
for the inclination angle from \citet{2006Sci...314...97K}, it is possible to use \rfr{angolo} and \rfr{uper} to constrain the spin's angles $\lambda_\mathrm{A},\,\eta_\mathrm{A}$; the allowed region in the plane $\grf{\lambda_\mathrm{A},\,\eta_\mathrm{A}}$ is displayed in Figure\,\ref{fig1}.
It turns out that
\begin{align}
85^\circ \lb{lam} &\lesssim \lambda_\mathrm{A}\lesssim 92^\circ, \\ \nonumber \\
266^\circ \lb{et}&\lesssim \eta_\mathrm{A} \lesssim 274^\circ.
\end{align}

\begin{figure}[H]
\centering
\centerline{
\vbox{
\begin{tabular}{c}
\epsfxsize= 16 cm\epsfbox{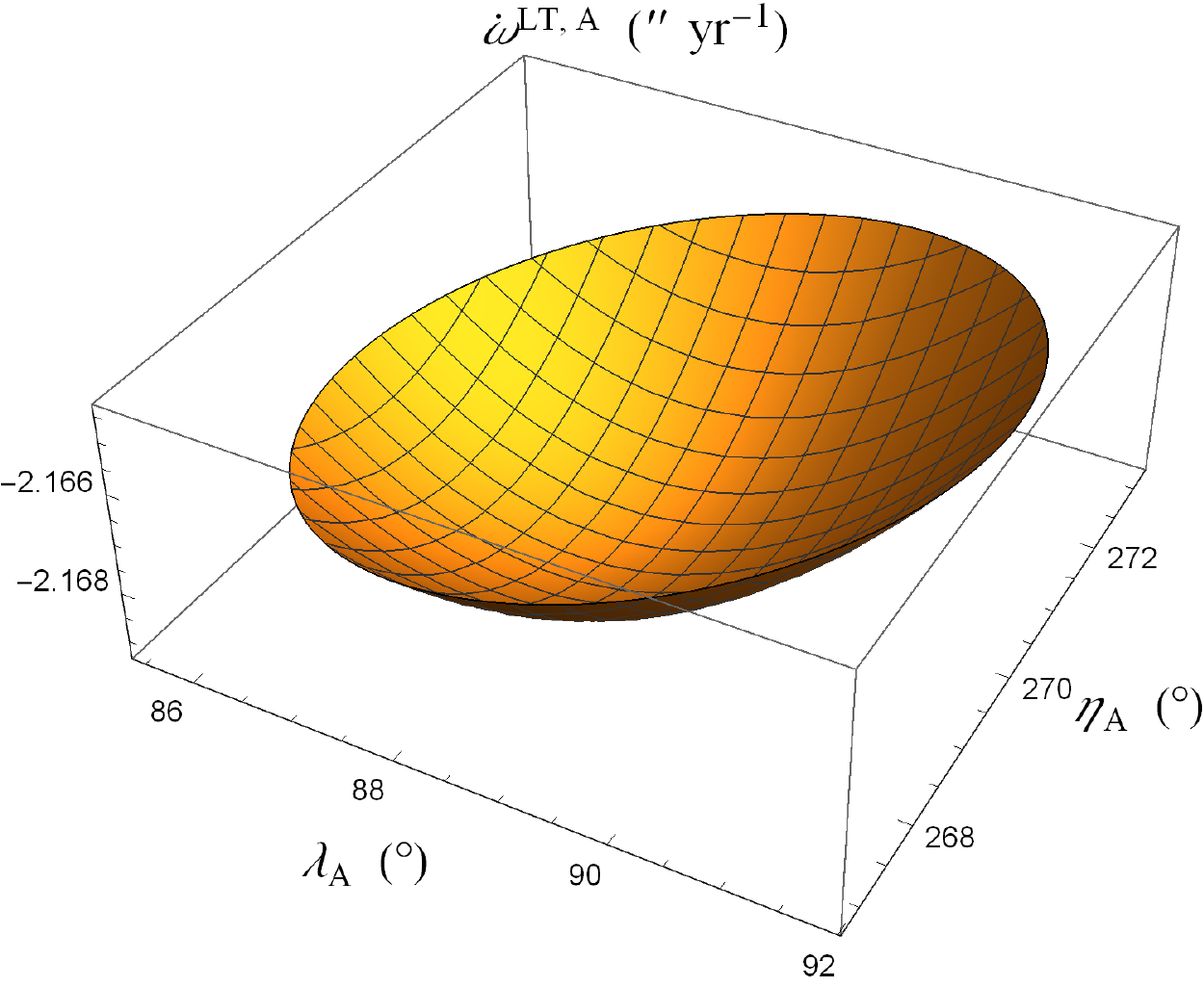}\\
\end{tabular}
}
}
\caption{
Plot of the Lense--Thirring periastron precession, in $\mathrm{''\,yr}^{-1}$, due to ${\bds S}^\mathrm{A}$ as a function of $\lambda_\mathrm{A},\,\eta_\mathrm{A}$ restricted to the allowed region of Figure\,\ref{fig1}. \Rfr{omeLT} was used along with \rfr{ine} and the system's values retrieved in \citet{2006Sci...314...97K}.
}\label{fig2}
\end{figure}

The Lense--Thirring periastron precession $\dot\omega^\mathrm{LT,\,A}$ due to ${\bds S}_\mathrm{A}$, calculated with \rfr{omeLT} and \rfr{ine}, is plotted in  Figure\,\ref{fig2}, in arcseconds per year $\ton{\mathrm{''\,yr}^{-1}}$, as a function of $\lambda_\mathrm{A},\,\eta_\mathrm{A}$ restricted to the allowed region of Figure\ref{fig1}.
It turns out that
\eqi
-2.16928\,\mathrm{''\,yr}^{-1}\leq\dot\omega^\mathrm{LT,\,A}\leq -2.16437\,\mathrm{''\,yr}^{-1},
\eqf
corresponding to a range variation of
\eqi
\Delta\dot\omega^\mathrm{LT\,A}\doteq\dot\omega^\mathrm{LT\,A}_\mathrm{max} - \dot\omega^\mathrm{LT\,A}_\mathrm{min} = 0.00491\,\mathrm{''\,yr}^{-1} = 8\times 10^{-8}\,\dot\omega^\mathrm{GR}.\lb{limitA}
\eqf
Figure 2 of \citet{2020MNRAS.497.3118H} tells that the experimental uncertainty  $\sigma_{\dot\omega_\mathrm{obs}}$ in determining the periastron precession may reach the level of \rfr{limitA} about after 2028, in the mid of the SKA 1-mid era.
\subsection{The contribution of ${\bds S}^\mathrm{B}$}\lb{sec1.2}
The contribution $\dot\omega^\mathrm{LT,\,B}$  due to ${\bds S}_\mathrm{B}$ to the overall Lense--Thirring periastron precession  can be obtained from \rfr{omeLT} with the exchange $\mathrm{A}\leftrightarrows\mathrm{B}$.
Since the spin period of \psrb\, \citep{2006Sci...314...97K}
\eqi
P_\mathrm{B} = 2.77\,\mathrm{s}
\eqf
is about 100 times longer than that of \psra,
\citet{2020MNRAS.497.3118H} neglected $\dot\omega^\mathrm{LT,\,B}$ in their analysis deeming it negligible.

\begin{figure}[H]
\centering
\centerline{
\vbox{
\begin{tabular}{c}
\epsfxsize= 16 cm\epsfbox{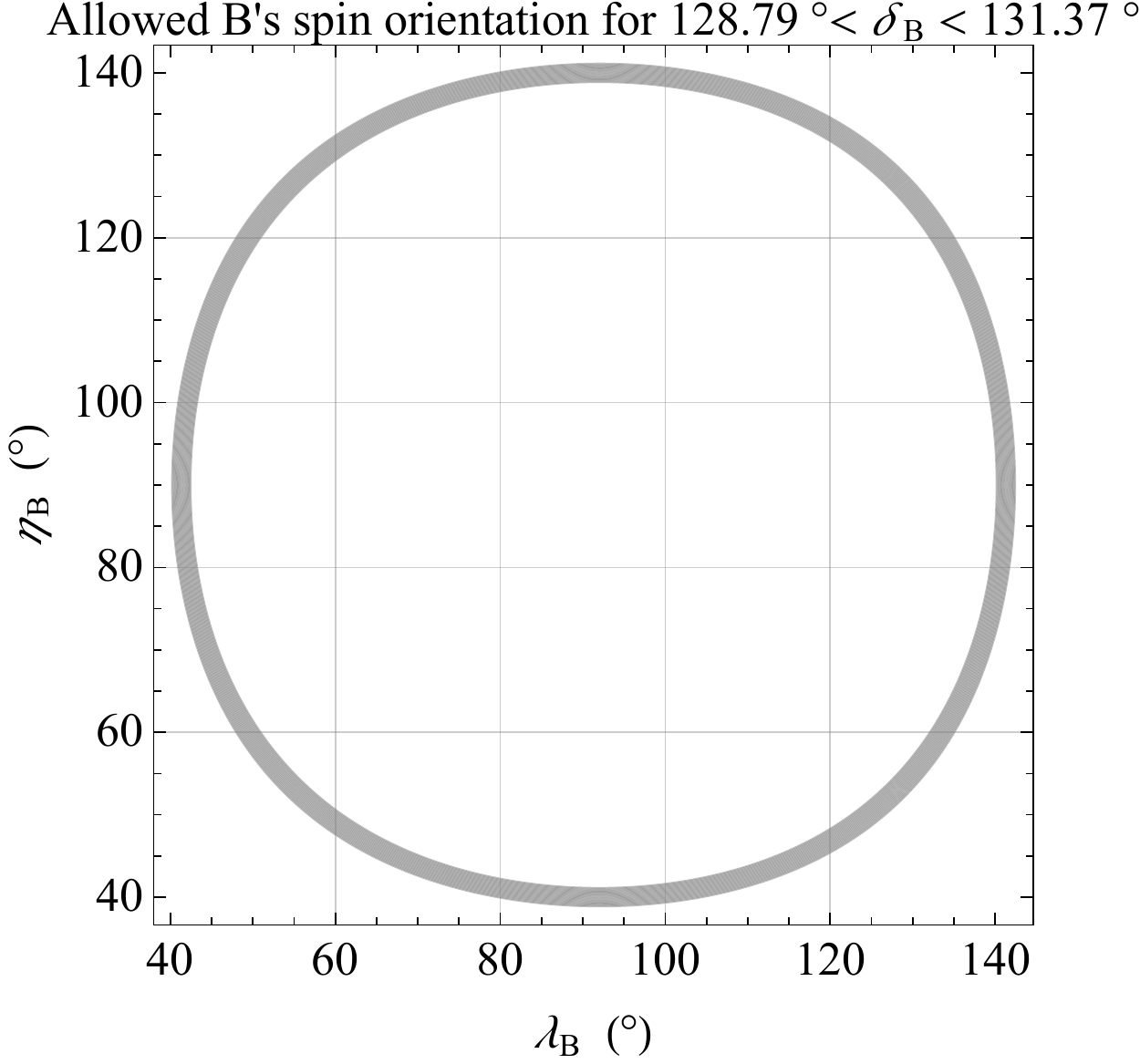}\\
\end{tabular}
}
}
\caption{
Allowed region, in the $\grf{\lambda_\mathrm{B},\,\eta_\mathrm{B}}$ plane, for the position angles $\lambda_\mathrm{B},\,\eta_\mathrm{B}$ of the spin ${\bds S}^\mathrm{B}$ of \psrb. It was obtained by plotting $\delta_\mathrm{B}$, calculated with $I=88.69^\circ$ \citep{2006Sci...314...97K}, as a function of  $\lambda_\mathrm{B},\,\eta_\mathrm{B}$ according to \rfr{angolo} with $\mathrm{A}\rightarrow\mathrm{B}$, and imposing the constraint of \rfr{uperB} determined for the epoch 2 May 2006 by \citet{2008Sci...321..104B}.}\label{fig3}
\end{figure}

\begin{figure}[H]
\centering
\centerline{
\vbox{
\begin{tabular}{c}
\epsfxsize= 16 cm\epsfbox{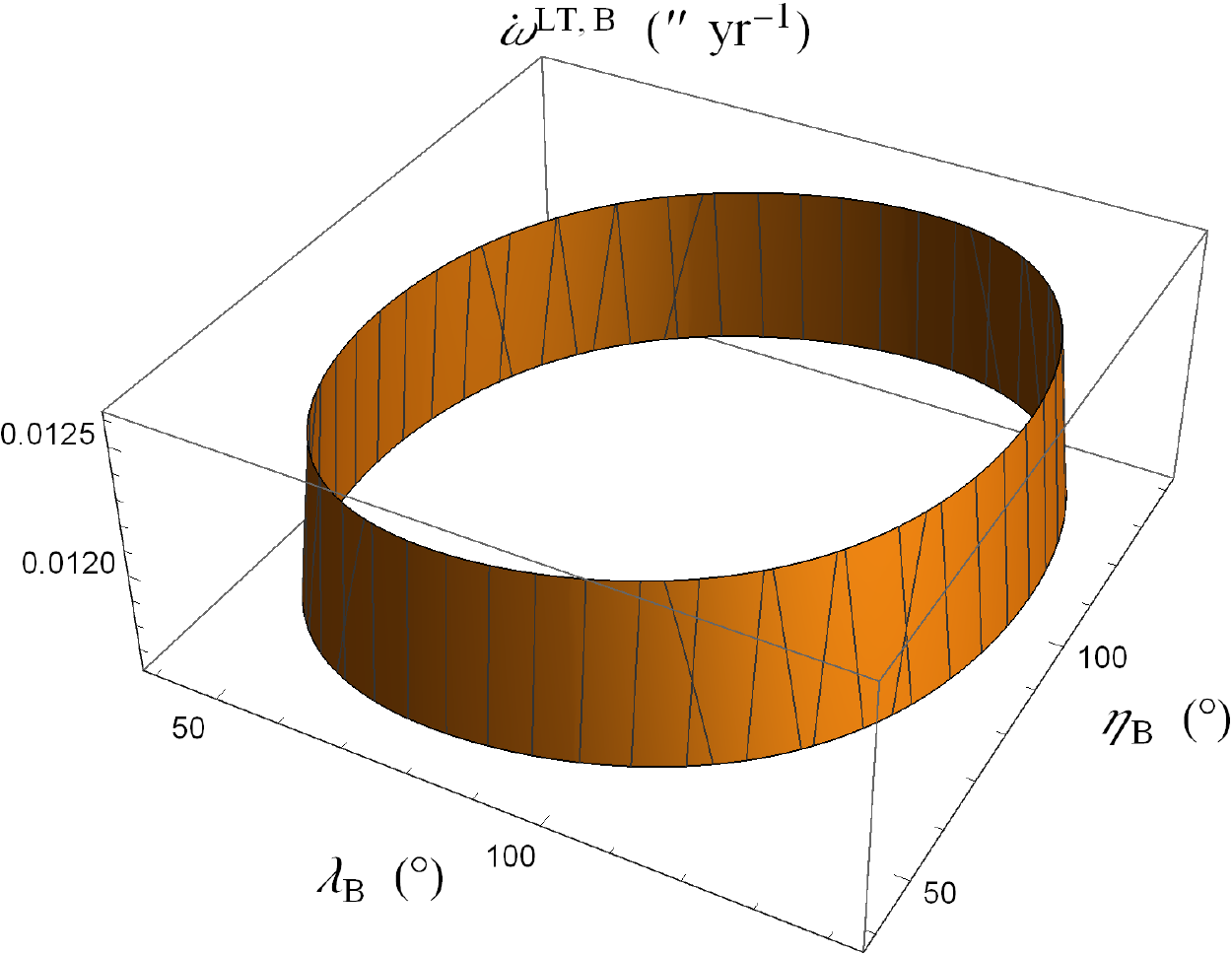}\\
\end{tabular}
}
}
\caption{
Plot of the Lense--Thirring periastron precession $\dot\omega^\mathrm{LT,\,B}$, in $\mathrm{''\,yr}^{-1}$, due to ${\bds S}^\mathrm{B}$ as a function of $\lambda_\mathrm{B},\,\eta_\mathrm{B}$ restricted to the allowed region of Figure\,\ref{fig3}. \Rfr{omeLT} was used with the replacement $\mathrm{A}\leftrightarrows\mathrm{B}$ along with \rfr{ine} and the system's values retrieved in \citet{2006Sci...314...97K}.
}\label{fig4}
\end{figure}

Here, it will be quantitatively assessed by relying upon the current knowledge of the spin-orbit geometry of \psrb.
The misalignment angle $\delta_\mathrm{B}$ between $\bds L$ and ${\bds S}^\mathrm{B}$ was accurately determined  for the epoch 2 May 2006 by \citet{2008Sci...321..104B} resulting equal to
\eqi
128.79^\circ\leq \delta_\mathrm{B}\leq 131.37^\circ\,\ton{99.7\%\,\mathrm{confidence}}.\lb{uperB}
\eqf
By using \rfr{angolo} written for $\mathrm{A}\rightarrow\mathrm{B}$, \rfr{uperB} permits to determine the allowed region in the $\grf{\lambda_\mathrm{B},\,\eta_\mathrm{B}}$ plane for the positional angles $\lambda_\mathrm{B},\,\eta_\mathrm{B}$ of ${\bds{\hat{S}}}^\mathrm{B}$. It is depicted in Figure\,\ref{fig3}.
It turns out to be a narrow annular region enclosed in a square $40^\circ\times 140^\circ$ wide.
Figure\,\ref{fig4} shows the corresponding $\dot\omega^\mathrm{LT,\,B}$ calculated for $\mathcal{I}_\mathrm{B} = \mathcal{I}_\mathrm{A}$.
It turns out that
\eqi
0.0116\,\mathrm{''\,yr}^{-1}\leq \dot\omega^\mathrm{LT,\,B}\leq 0.0126\,\mathrm{''\,yr}^{-1},
\eqf
corresponding to
\eqi
\dot\omega^\mathrm{LT,\,B}\simeq 2\times 10^{-7}\,\dot\omega^\mathrm{GR}.\lb{limitB}
\eqf
According to Figure 2 of \citet{2020MNRAS.497.3118H},  $\sigma_{\dot\omega_\mathrm{obs}}$  may reach the level of \rfr{limitB} between the end of the MeerKAT+ era and the beginning of the SKA 1-mid phase about around  2025.
%
%
%
%
%
\section{The spin-orbit precessions of the inclination and the node}\lb{sec2}
According to \citet{2021MNRAS.504.2094K}, \psrab\, is one of the few binary pulsars for which $I$ and $\Omega$ can be uniquely determined by means of long term monitoring of the interstellar scintillation following a strategy put forth for the first time by \citet{1984Natur.310..300L}; see, e.g., the determinations of $I$ and\footnote{\citet{2014ApJ...787..161R} used a different convention for the node, which was accounted for by \citet[][p.\,3121]{2020MNRAS.497.3118H} in releasing their result for $\Omega$.}  $\Omega$ by \citet{2014ApJ...787..161R}.
Nonetheless, further investigations are required to assess if the results from the scintillation method are consistent with those obtained from the standard pulsar timing including the Rotating Vector Model (RVM) \citep{2021MNRAS.504.2094K}.

Thus, the gravitomagnetic spin-orbit rates of $I$ and $\Omega$ are accurately calculated in order to check if a possible future measurement of them may be viable. It is important to stress that, contrary to the periastron, neither the inclination nor the node undergo  1pN$+$2pN mass-dependent gravitoelectric shifts. Their full spin-orbit rates can be explicitly expressed as \citep{2017EPJC...77..439I}
\begin{align}
\dot I^\mathrm{LT} \nonumber \lb{InclLT}& = \rp{2\,G\,S^\mathrm{A}}{c^2\,a^3\,\ton{1-e^2}^{3/2}}\ton{1+\rp{3}{4}\rp{M_\mathrm{B}}{M_\mathrm{A}}}{\bds{\hat{S}}}^\mathrm{A}\bds\cdot\bds l + \mathrm{A}\leftrightarrows\mathrm{B} = \\ \nonumber \\
& =\rp{2\,G\,S^\mathrm{A}}{c^2\,a^3\,\ton{1-e^2}^{3/2}}\ton{1+\rp{3}{4}\rp{M_\mathrm{B}}{M_\mathrm{A}}}\cos\eta_\mathrm{A}\,\sin\lambda_\mathrm{A} + \mathrm{A}\leftrightarrows\mathrm{B}, \\ \nonumber \\
\dot \Omega^\mathrm{LT} \nonumber \lb{NodLT}& = \rp{2\,G\,S^\mathrm{A}}{c^2\,a^3\,\sin I\,\ton{1-e^2}^{3/2}}\ton{1+\rp{3}{4}\rp{M_\mathrm{B}}{M_\mathrm{A}}}{\bds{\hat{S}}}^\mathrm{A}\bds\cdot\bds m + \mathrm{A}\leftrightarrows\mathrm{B} =\\ \nonumber \\
& =\rp{2\,G\,S^\mathrm{A}}{c^2\,a^3\,\ton{1-e^2}^{3/2}}\ton{1+\rp{3}{4}\rp{M_\mathrm{B}}{M_\mathrm{A}}}\ton{\cos\lambda_\mathrm{A} + \cot I\,\sin\eta_\mathrm{A}\,\sin\lambda_\mathrm{A}} + \mathrm{A}\leftrightarrows\mathrm{B}.
\end{align}
For more or less general expressions of the two-body Lense--Thirring out--of--plane orbital precessions, see also \citet{Micha60}, \citet[][Equations\,(5.11b)-(511c) ]{1988NCimB.101..127D}, \citet[][Section\,4.4.4]{1991ercm.book.....B} and \citet[][Equation\,(3.27)]{1992PhRvD..45.1840D}. Moreover, \citet{1975PhRvD..12..329B} expressed the two-body Lense--Thirring orbital precessions in vectorial form.

\begin{figure}[H]
\centering
\centerline{
\vbox{
\begin{tabular}{cc}
\epsfxsize= 7.5 cm\epsfbox{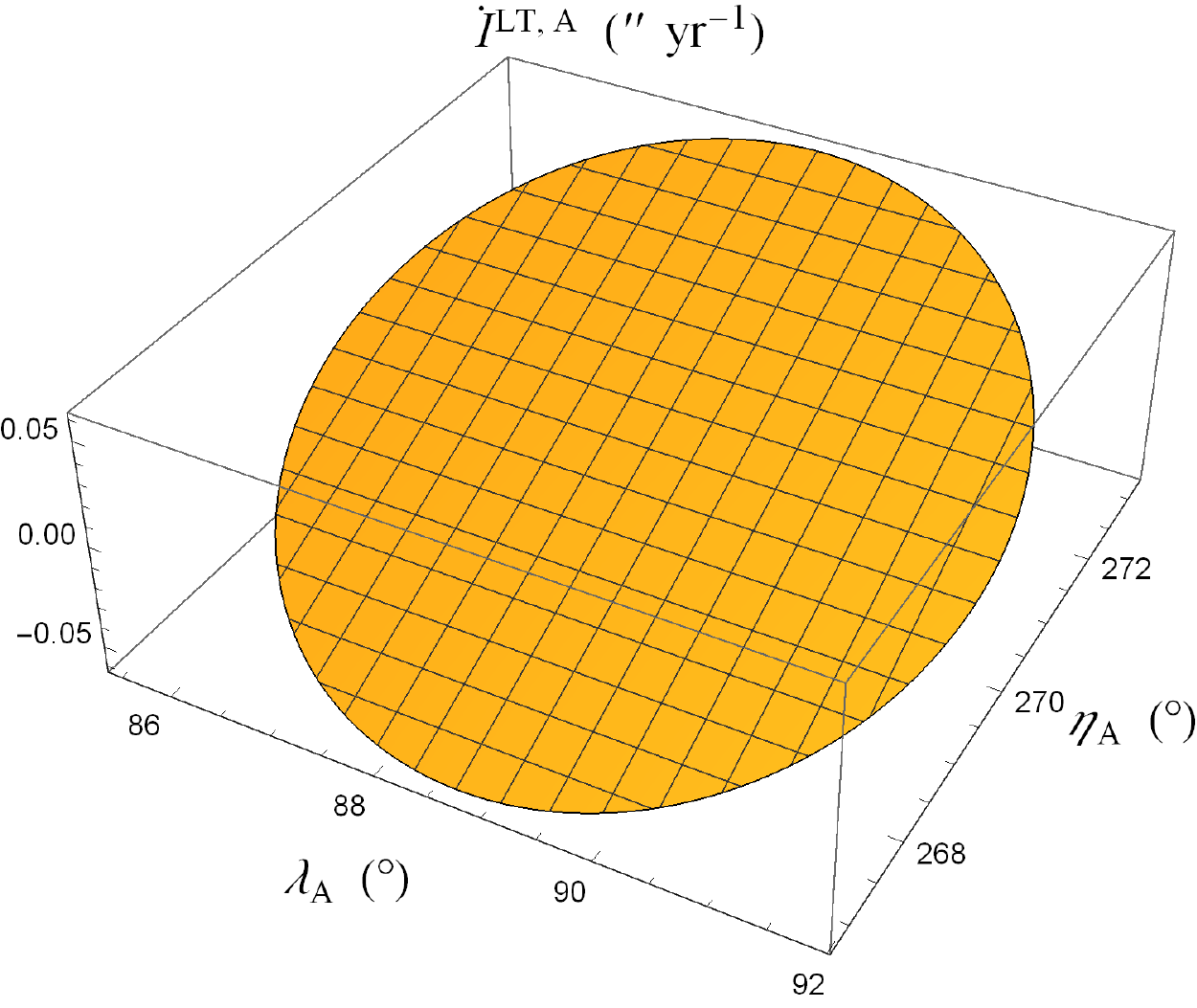} & \epsfxsize= 7.5 cm\epsfbox{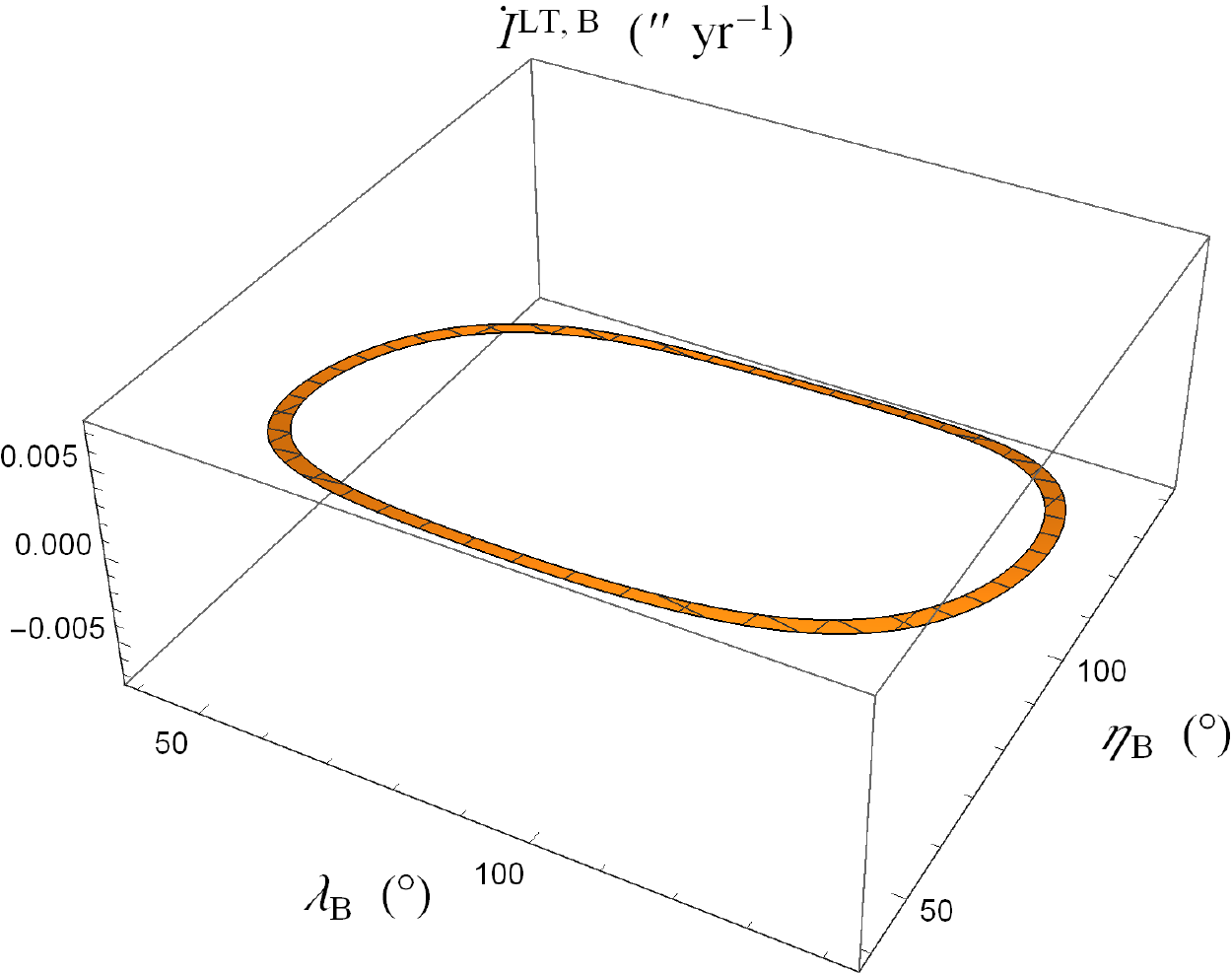}\\
\epsfxsize= 7.5 cm\epsfbox{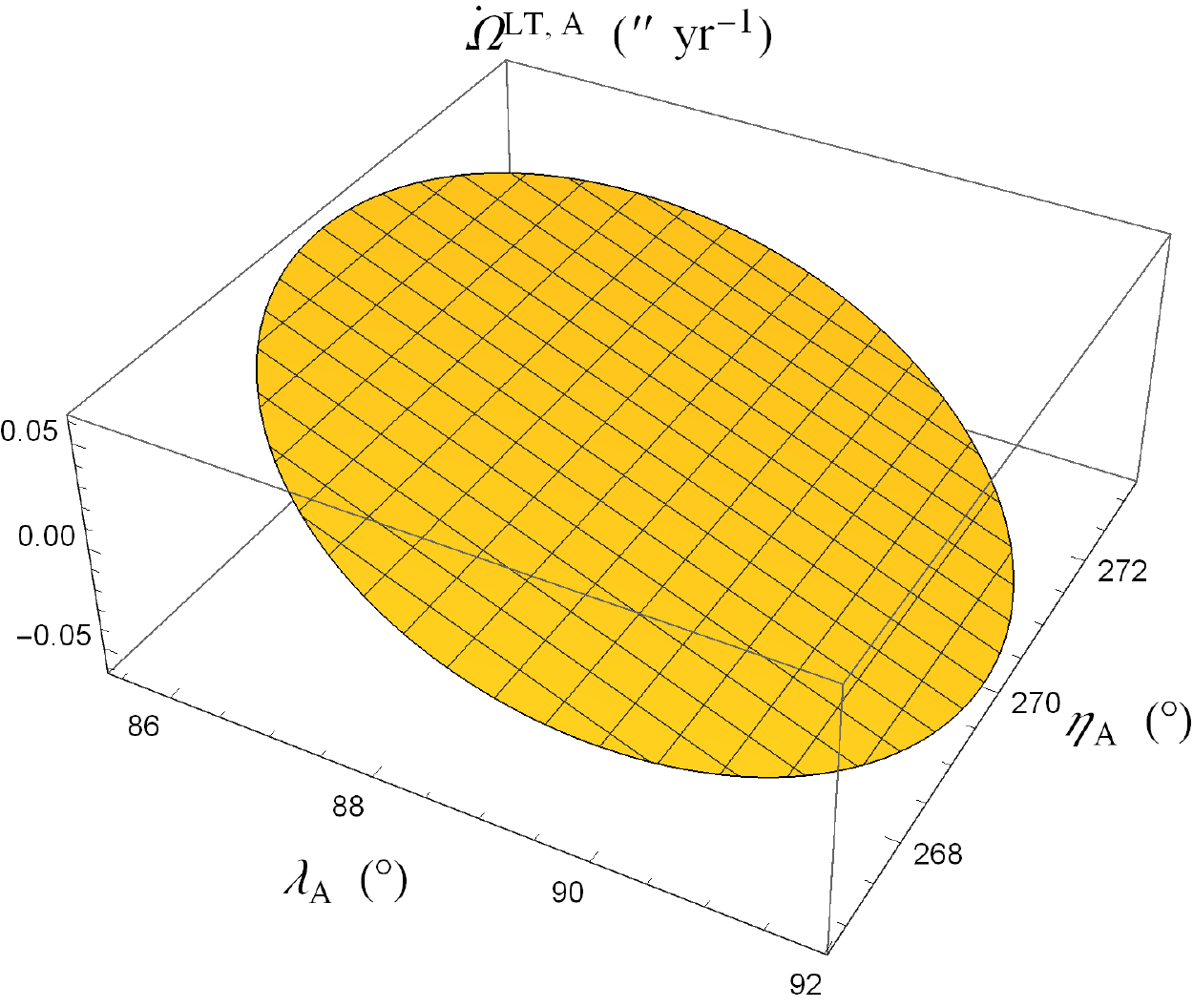} & \epsfxsize= 7.5 cm\epsfbox{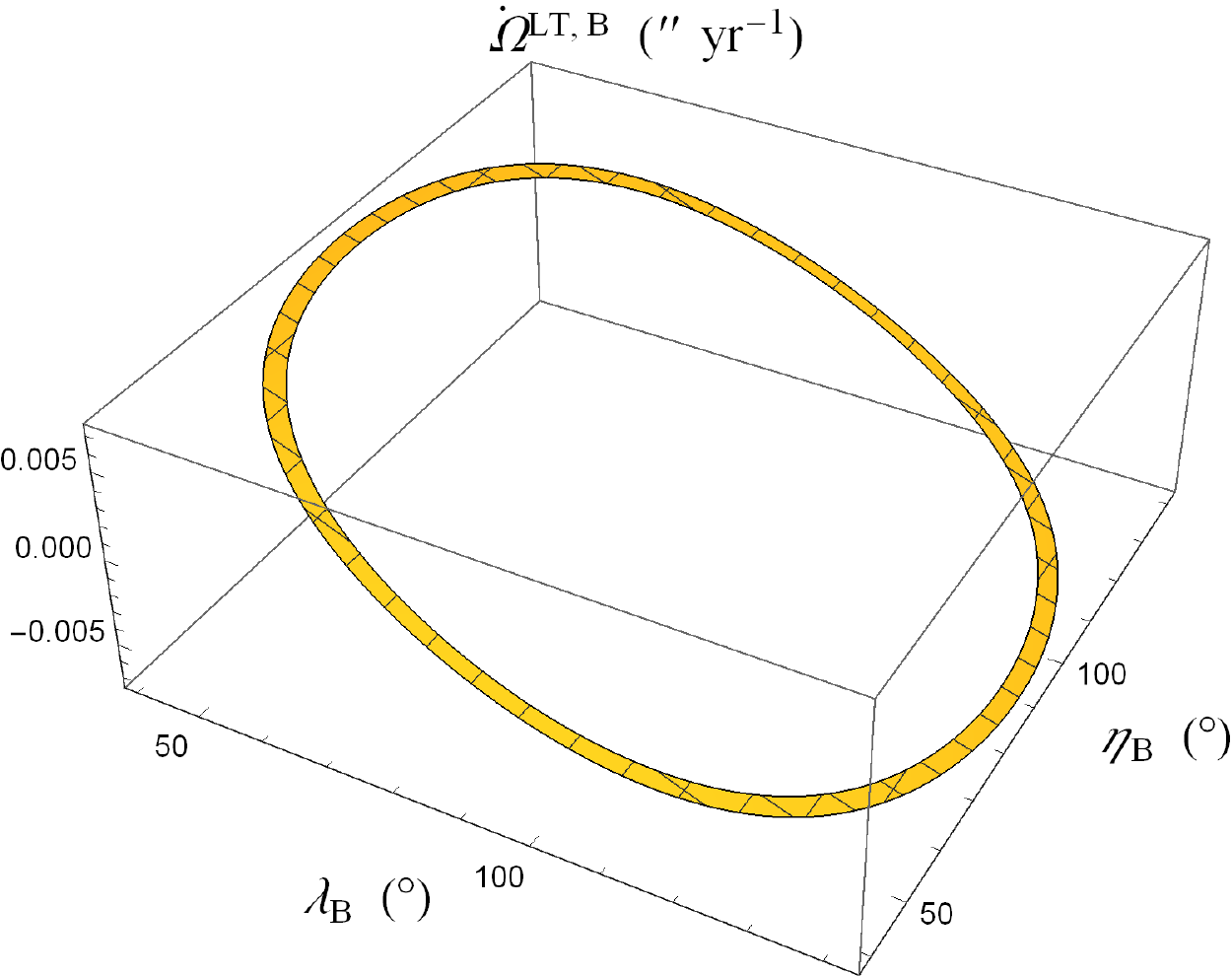}\\
\end{tabular}
}
}
\caption{
Plot of the Lense--Thirring inclination and node precessions $\dot I^\mathrm{LT,\,A},\,\dot I^\mathrm{LT,\,B},\,\dot \Omega^\mathrm{LT,\,A},\,\dot \Omega^\mathrm{LT,\,B}$, in $\mathrm{''\,yr}^{-1}$, due to ${\bds S}^\mathrm{A}$ and ${\bds S}^\mathrm{B}$ as  functions of $\lambda_\mathrm{A},\,\eta_\mathrm{A}$ and $\lambda_\mathrm{B},\,\eta_\mathrm{B}$, respectively, restricted to the allowed regions of Figure\,\ref{fig1} and Figure\,\ref{fig3}. \Rfrs{InclLT}{NodLT} were used  along with \rfr{ine} for the moments of inertia and the system's values retrieved in \citet{2006Sci...314...97K}.
}\label{fig5}
\end{figure}

Figure\,\ref{fig5} displays the Lense--Thirring precessions $\dot I^\mathrm{LT,\,A},\,\dot I^\mathrm{LT,\,B},\,\dot \Omega^\mathrm{LT,\,A},\,\dot \Omega^\mathrm{LT,\,B}$ calculated by means of \rfrs{InclLT}{NodLT} within the allowed regions for ${\bds{\hat{S}}}^\mathrm{A}$ and ${\bds{\hat{S}}}^\mathrm{B}$, respectively, in the planes $\grf{\lambda_\mathrm{A},\,\eta_\mathrm{A}}$ and $\grf{\lambda_\mathrm{B},\,\eta_\mathrm{B}}$.
It turns out that the maximum size of the absolute value of the precessions due to the spin of \psra\, is
\eqi
\left|\dot I^\mathrm{LT,\,A}\right|\simeq \left|\dot \Omega^\mathrm{LT,\,A}\right|\lesssim 0.05\,\mathrm{''\,yr}^{-1},\lb{Ahia}
\eqf
while for the rates induced by \psrb\, the upper bound is
\eqi
\left|\dot I^\mathrm{LT,\,B}\right|\simeq \left|\dot \Omega^\mathrm{LT,\,B}\right|\lesssim 0.005\,\mathrm{''\,yr}^{-1}.\lb{Bhia}
\eqf
Furthermore, for some particular orientations of ${\bds{\hat{S}}}^\mathrm{A},\,{\bds{\hat{S}}}^\mathrm{B}$, they can even vanish.
The current experimental errors in determining $I$ and $\Omega$ with the scintillation technique are several orders of magnitude larger than \rfrs{Ahia}{Bhia}; suffice it to say that \citet{2014ApJ...787..161R} report
\begin{align}
\sigma_{I_\mathrm{obs}} &= 0.5^\circ, \\ \nonumber \\
\sigma_{\Omega_\mathrm{obs}} &= 2^\circ
\end{align}
over a time span of $1.5$ yr.

\begin{figure}[H]
\centering
\centerline{
\vbox{
\begin{tabular}{c}
\epsfxsize= 16 cm\epsfbox{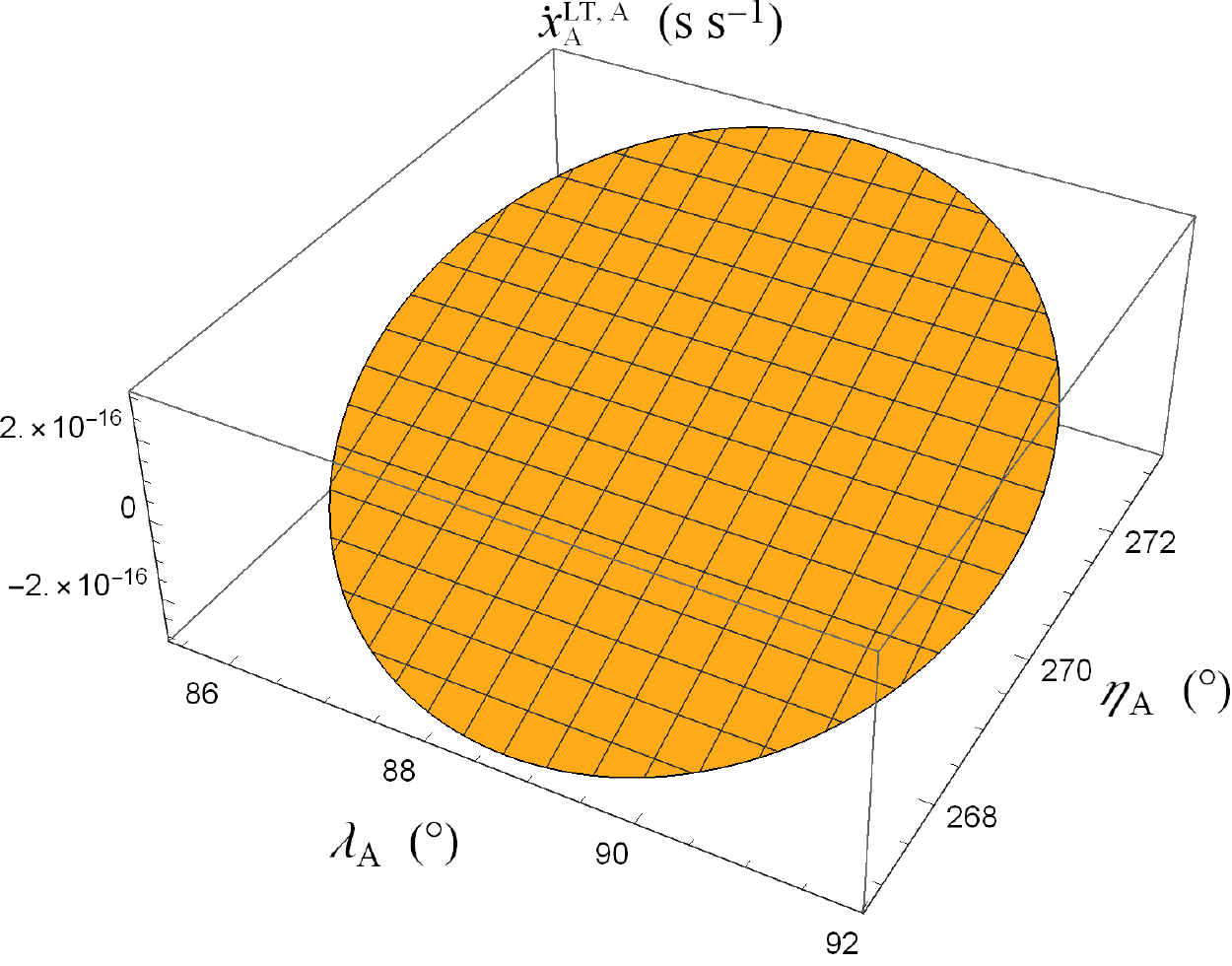}\\
\end{tabular}
}
}
\caption{
Plot of the Lense--Thirring rate of change $\dot x_\mathrm{A}^\mathrm{LT,\,A}$ of the projected semimajor axis $x_\mathrm{A}$ of \psra, in $\mathrm{s\,s}^{-1}$, due to ${\bds S}^\mathrm{A}$ as a function of $\lambda_\mathrm{A},\,\eta_\mathrm{A}$ restricted to the allowed region of Figure\,\ref{fig1}. \Rfr{InclLT} was used  along with \rfr{ine} and the system's values retrieved in \citet{2006Sci...314...97K}.
}\label{fig6}
\end{figure}

As far as the gravitomagnetic spin-orbit precession of the inclination is concerned, another possibility to measure it consists, in principle, in determining the rate of change
\eqi
\dot x_\mathrm{A} = x_\mathrm{A}\cot I\,\dot I = \rp{a_\mathrm{A}\,\cos I}{c}\,\dot I
\eqf
of the projected semimajor axis $x_\mathrm{A}$ of \psra.
Figure\,\ref{fig6} shows $\dot x_\mathrm{A}^\mathrm{LT,\,A}$ as a function of $\lambda_\mathrm{A},\,\eta_\mathrm{A}$ restricted to their allowed region of Figure\,\ref{fig1}.
The resulting maximum value of the size of the Lense--Thirring rate of change $\dot x_\mathrm{A}^\mathrm{LT,\,A}$ of the projected semimajor axis of \psra\, is
\eqi
\left|\dot x_\mathrm{A}^\mathrm{LT,\,A}\right|\lesssim 2\times 10^{-16}\,\mathrm{s\,s}^{-1}.\lb{megax}
\eqf
Note that, for certain values of $\lambda_\mathrm{A},\,\eta_\mathrm{A}$, $\dot x^\mathrm{LT,\,A}_\mathrm{A}$ vanishes.
Despite, to date,  measurements of $\dot x_\mathrm{A}$ are seemingly missing in the literature, a guess on the potentially achievable accuracy in its determination may be inferred from the published results. From \citep{2006Sci...314...97K}
\eqi
\sigma_{x_\mathrm{A}^\mathrm{obs}} = 1\times 10^{-6}\,\mathrm{s}
\eqf
over a time span of \citep{2006Sci...314...97K}
\eqi
\Delta t = 2.5\,\mathrm{yr} =7.88\times 10^7\,\mathrm{s},
\eqf
a plausible estimate of the accuracy in determining $\dot x_\mathrm{A}$ can be tentatively argued, amounting to
\eqi
\sigma_{\dot x_\mathrm{A}^\mathrm{obs}} \simeq 1.2\times 10^{-14}\,\mathrm{s\,s}^{-1}.\lb{errx}
\eqf
It should be remarked that \rfr{errx} is just two orders of magnitude larger than \rfr{megax}; thus, it does not seem unreasonable to expect that a measurement of $\dot x_\mathrm{A}^\mathrm{LT,\,A}$ may become feasible in the near future.

A binary system for which the Lense--Thirring rate of the projected semimajor axis seems to be measurable in the forthcoming years is PSR J1757--1854 \citep{2018MNRAS.475L..57C}.
For its pulsar p, by assuming \citep{2005ApJ...629..979L}
\eqi
\mathcal{I}_\mathrm{p} = 1.2\times 10^{38}\,\mathrm{kg\,m}^2,
\eqf
one has \citep{2018MNRAS.475L..57C}
\eqi
\dot x^\mathrm{LT}_\mathrm{p}\simeq 1.9\times 10^{-14}\,\mathrm{s\,s}^{-1}.\lb{bib}
\eqf
\citet{2018MNRAS.475L..57C} reported an experimental uncertainty in determining $x_\mathrm{p}$
\eqi
\sigma_{x_\mathrm{p}} = 5\times 10^{-6}\,\mathrm{s}
\eqf
over an observational time span
\eqi
\Delta t = 1.6\,\mathrm{yr} = 5.05\times 10^7\,\mathrm{s},
\eqf
corresponding to a possible error in measuring $\dot x_\mathrm{p}$ of the order of
\eqi
\sigma_{\dot x_\mathrm{p}} \simeq 9.9\times 10^{-14}\,\mathrm{s\,s}^{-1}.
\eqf
Probably relying upon an analogous guess, \citet{2018MNRAS.475L..57C} concluded that a  measurement of \rfr{bib} to within $3\sigma$ would be possible in the following 8-9 years.
\section{The quadrupole-induced orbital precessions}\lb{sec3}
In principle, for an arbitrary orientation in space of ${\bds{\hat{S}}}^\mathrm{A}$ and ${\bds{\hat{S}}}^\mathrm{B}$, $I,\,\Omega$ and $\omega$ undergo Newtonian long-term rates of change due to the quadrupole mass moments $Q^\mathrm{A},\,Q^\mathrm{B}$ of A and B. Their explicit general expressions are
\citep{2017EPJC...77..439I}
\begin{align}
\dot I^Q \lb{iQ}& = \rp{3\,\nk\,Q_\mathrm{A}}{2\,a^2\,\ton{1-e^2}^2\,M_\mathrm{A}}\,\ton{{\bds{\hat{S}}}^\mathrm{A}\bds\cdot\bds l}\,\ton{{\bds{\hat{S}}}^\mathrm{A}\bds\cdot\bds n}  + \mathrm{A}\leftrightarrows\mathrm{B} = \\ \nonumber \\
&= \rp{3\,\nk\,Q_\mathrm{A}}{2\,a^2\,\ton{1-e^2}^2\,M_\mathrm{A}}\,\cos\eta_\mathrm{A}\,\sin\lambda_\mathrm{A}\,\ton{\cos I\, \cos\lambda_\mathrm{A}- \sin I\,\sin\eta_\mathrm{A}\,\sin\lambda_\mathrm{A}} + \mathrm{A}\leftrightarrows\mathrm{B}, \\ \nonumber \\
\dot\Omega^Q   \lb{NQ} \nonumber & =  \rp{3\,\nk\,Q_\mathrm{A}\,\csc I}{2\,a^2\,\ton{1-e^2}^2\,M_\mathrm{A}}\,\ton{{\bds{\hat{S}}}^\mathrm{A}\bds\cdot\bds m}\,\ton{{\bds{\hat{S}}}^\mathrm{A}\bds\cdot\bds n}  + \mathrm{A}\leftrightarrows\mathrm{B} = \\ \nonumber \\
\nonumber &= \rp{3\,\nk\,Q_\mathrm{A}}{4\,a^2\,\ton{1-e^2}^2\,M_\mathrm{A}}\,\qua{2 \cos I\,\ton{\cos^2\lambda_\mathrm{A} - \sin^2\eta_\mathrm{A}\,\sin^2\lambda_\mathrm{A}} + \right.\\ \nonumber \\
&+\left. \cos 2I\,\csc I\,\sin\eta_\mathrm{A}\,\sin 2\lambda_\mathrm{A}} + \mathrm{A}\leftrightarrows\mathrm{B}, \\ \nonumber \\
\dot\omega^Q   \lb{oQ} \nonumber & =  -\rp{3\,\nk\,Q_\mathrm{A}}{4\,a^2\,\ton{1-e^2}^2\,M_\mathrm{A}}\,\grf{2 - 3\qua{\ton{{\bds{\hat{S}}}^\mathrm{A}\bds\cdot\bds l}^2 + \ton{{\bds{\hat{S}}}^\mathrm{A}\bds\cdot\bds m}^2} + \right.\\ \nonumber \\
\nonumber &+\left. 2 \cot I\,\ton{{\bds{\hat{S}}}^\mathrm{A}\bds\cdot\bds m}\,\ton{{\bds{\hat{S}}}^\mathrm{A}\bds\cdot\bds n} }  + \mathrm{A}\leftrightarrows\mathrm{B} = \\ \nonumber \\
\nonumber &= \rp{3\,\nk\,Q_\mathrm{A}}{32\,a^2\,\ton{1-e^2}^2\,M_\mathrm{A}}\,\qua{-\ton{3 + 5 \cos 2I}\,\ton{1 + 3 \cos 2\lambda_\mathrm{A}} + \right.\\ \nonumber \\
\nonumber & +\left. 2\,\ton{1 - 5 \cos 2I}\,\cos 2\eta_\mathrm{A}\,\sin^2\lambda_\mathrm{A} + \right.\\ \nonumber \\
&+\left. 2\,\ton{\cos I\,- 5 \cos 3I}\,\csc I\,\sin\eta_\mathrm{A}\,\sin 2\lambda_\mathrm{A}} + \mathrm{A}\leftrightarrows\mathrm{B}.
\end{align}
See also \citet{1975PhRvD..12..329B} for their effects in vectorial form. References like, e.g., \citet{1976ApJ...207..574S,1998MNRAS.298...67W} dealing with the orbital elements are either valid only for the restricted two-body case or are restricted to some particular spin-orbit geometry.

\citet{2020MNRAS.497.3118H}, relying upon the relations by \citet{2013ApJ...777...68B} to calculate $Q^\mathrm{A}$, claimed that
\eqi
\dot\omega^{Q,\,\mathrm{A}} \simeq 0.0001\,\mathrm{''\,yr}^{-1}= 1.7\times 10^{-9}\,\dot\omega^\mathrm{GR},\lb{bau}
\eqf
which would be likely negligible in the foreseeable future. Here, an independent evaluation will be offered in light of \rfrs{iQ}{oQ} and of the most recent results in determining the key physical parameters of neutron stars in a rather EOS-independent way \citep{2021PhRvL.126r1101S}.

According to \citet{2021PhRvL.126r1101S}, the dimensional quadrupole mass moment
of a neutron star like \psra\, can be expressed as
\eqi
Q_\mathrm{A} = -\overline{Q}_\mathrm{A}\,\rp{M_\mathrm{A}^3\,G^2\,\chi_\mathrm{A}^2}{c^4},\lb{Qu}
\eqf
where $\overline{Q}_\mathrm{A}$ and $\chi_\mathrm{A}$ are dimensionless parameters related to the MOI; \rfr{Qu} is rather insensitive to the EOS \citep{2021PhRvL.126r1101S}. In particular,
\eqi
\chi_\mathrm{A} \doteq \rp{2\uppi\nu_\mathrm{A}\,M_\mathrm{A}\,G\,\overline{\mathcal{I}}_\mathrm{A}}{c^3},\lb{chi}
\eqf
where the dimensionless MOI-type parameter is defined as
\eqi
\overline{\mathcal{I}}_\mathrm{A} = \rp{c^4\,\mathcal{I}_\mathrm{A}}{M_\mathrm{A}^3\,G^2}.\lb{Ibar}
\eqf
Inserting \rfrs{chi}{Ibar} in \rfr{Qu}, the quadrupole mass moment of \psra\, can be written as
\eqi
Q_\mathrm{A} = - \overline{Q}_\mathrm{A}\,\rp{4\uppi^2\,\nu^2_\mathrm{A}\,\mathcal{I}^2_\mathrm{A}}{c^2\,M_\mathrm{A}}.\lb{QU}
\eqf
\citet{2021PhRvL.126r1101S} provided estimates of the moment of inertia $\mathcal{I}_\mathrm{A}$ of \psra, and of the dimensionless parameter
\eqi
\overline{Q}_\star\simeq 6\lb{QUBAR}
\eqf
for the isolated millisecond pulsar PSR J0030+0451 whose spin frequency $\nu_\star = 205.53\,\mathrm{Hz}$ is about five times higher than that of \psra. By calculating \rfr{QU} for \psra, it is possible to obtain
\eqi
Q_\mathrm{A}= -\overline{Q}_\mathrm{A}\,8\times 10^{33}\,\mathrm{kg\,m}^2.\lb{QUA}
\eqf
\Rfr{QUA} is comparable to the value calculated by \citet[][Equations\,(88)-(89)]{2017EPJC...77..439I} relying upon \citet{1999ApJ...512..282L}.

Incidentally, \rfr{QU} yields for  \psrb\, a value four orders of magnitude smaller than \rfr{QUA}.

By  calculating \rfr{oQ} with, say, \rfr{QUBAR} in \rfr{QUA}, its plot in the allowed region of \rfrs{lam}{et} for $\lambda_\mathrm{A},\,\eta_\mathrm{A}$,
returns
\eqi
\dot\omega^{Q,\,\mathrm{A}}\simeq 0.00017\,\mathrm{''\,yr}^{-1},
\eqf
in agreement  with \rfr{bau}.

As far as the other orbital elements are concerned, it turns out that
\begin{align}
\left|\dot I^{Q,\,\mathrm{A}}\right| &\lesssim 5\times 10^{-6}\,\mathrm{''\,yr}^{-1}, \\ \nonumber \\
\left|\dot\Omega^{Q,\,\mathrm{A}}\right| &\lesssim 5\times 10^{-6}\,\mathrm{''\,yr}^{-1}, \\ \nonumber \\
\left|\dot x_\mathrm{A}^{Q,\,\mathrm{A}}\right| &\lesssim 4\times 10^{-20}\,\mathrm{s\,s}^{-1}.
\end{align}
Such so small figures are likely negligible, from an observational point of view, in any foreseeable future.
\section{Summary and conclusions}\lb{sec4}
The expected improvements in the experimental accuracy $\sigma_{\dot\omega_\mathrm{obs}}$ in determining the periastron precession $\dot\omega$ of \psrab, which, in units of the general relativistic mass-only gravitoelectric precession $\dot\omega^\mathrm{GR}$, may reach the $\simeq 10^{-7}-10^{-8}\,\dot\omega^\mathrm{GR}$ level by 2030 thanks to the SKA 1-mid facility, requires a careful modeling of all the dynamical contributions to the long-term rate of change of such an orbital element.

Among them, there is the general relativistic gravitomagnetic Lense--Thirring periastron rate $\dot\omega^\mathrm{LT}$, that is dominated by the contribution $\dot\omega^\mathrm{LT,\,A}$ induced by the spin angular momentum ${\bds S}^\mathrm{A}$ of \psra, assumed aligned with the orbital angular momentum $\bds L$. It is particularly important since it depends on the pulsar's moment of inertia $\mathcal{I}_\mathrm{A}$ whose determination can give valuable information on the equation of state of the ultradense matter inside neutron stars.

Here, it was demonstrated that neglecting the part of $\dot\omega^\mathrm{LT,\,A}$ due to the misalignment between ${\bds S}^\mathrm{A}$ and $\bds L$, which corresponds to an allowed orientation in space of ${\bds{\hat{S}}}^\mathrm{A}$ constrained within
$85^\circ \lesssim \lambda_\mathrm{A} \lesssim 92^\circ,\,266^\circ \lesssim \eta_\mathrm{A} \lesssim 274^\circ$,
would introduce an error which may be as large as $\Delta\dot\omega^\mathrm{LT,\,A}\simeq 8\times 10^{-8}\,\dot\omega^\mathrm{GR}$. Furthermore, also the contribution $\dot\omega^\mathrm{LT,\,B}$ to $\dot\omega^\mathrm{LT}$ from the spin ${\bds S}^\mathrm{B}$ of \psrb, always neglected in all the analyses published so far, should be taken into account since its magnitude may be as large as $\dot\omega^\mathrm{LT,\,B}\simeq 2\times 10^{-7}\,\dot\omega^\mathrm{GR}$.

Also the orbital inclination $I$ and the node $\Omega$  undergo Lense--Thirring precessions whose sizes, for \psrab, are of the order of  $\left|\dot I^\mathrm{LT}\right|\simeq\left|\dot\Omega^\mathrm{LT}\right|\lesssim 0.05\,\mathrm{arcseconds\,per\,year}$. The current experimental uncertainty in measuring $I$ and $\Omega$ of \psrab\, with the scintillation technique is several orders of magnitude larger. Should the rate $\dot x_\mathrm{A}$ of the projected semimajor axis $x_\mathrm{A}$ of \psra\, be measurable, the size of its expected gravitomagnetic spin-orbit contribution is $\left|\dot x^\mathrm{LT}_\mathrm{A}\right|\lesssim 2\times 10^{-16}\,\mathrm{s\,s}^{-1}$. The experimental uncertainty $\sigma_{\dot x_\mathrm{A}^\mathrm{obs}}$ in measuring $\dot x_\mathrm{A}$, hypothesized on the basis of the data released in 2006, is of the order of $\simeq 10^{-14}\,\mathrm{s\,s}^{-1}$.

The contribution to $\dot\omega$ by the quadrupole mass moment $Q^\mathrm{A}$ of \psra, is confirmed to be likely negligible in the foreseeable future since it is as little as $\dot\omega^{Q,\,A}\simeq 10^{-9}\,\dot\omega^\mathrm{GR}$.
%
%
\section*{Data availability}
No new data were generated or analysed in support of this research.
\bibliography{psrbib}{}

\begin{thebibliography}{57}
\expandafter\ifx\csname natexlab\endcsname\relax\def\natexlab#1{#1}\fi

\bibitem[{{Barker} \& {O'Connell}(1975)}]{1975PhRvD..12..329B}
{Barker} B.~M., {O'Connell} R.~F., 1975, Phys. Rev. D, 12, 329

\bibitem[{{Baub{\"o}ck} {et~al}\mbox{.}(2013){Baub{\"o}ck}, {Berti}, {Psaltis},
  \& {{\"O}zel}}]{2013ApJ...777...68B}
{Baub{\"o}ck} M., {Berti} E., {Psaltis} D., {{\"O}zel} F., 2013, \apj, 777, 68

\bibitem[{{Breton} {et~al}\mbox{.}(2008){Breton}, {Kaspi}, {Kramer},
  {McLaughlin}, {Lyutikov}, {Ransom}, {Stairs}, {Ferdman}, {Camilo}, \&
  {Possenti}}]{2008Sci...321..104B}
{Breton} R.~P. {et~al.}, 2008, Science, 321, 104

\bibitem[{{Brumberg}(1991)}]{1991ercm.book.....B}
{Brumberg} V.~A., 1991, {Essential Relativistic Celestial Mechanics}. Adam
  Hilger, Bristol

\bibitem[{{Burgay} {et~al}\mbox{.}(2003){Burgay}, {D'Amico}, {Possenti},
  {Manchester}, {Lyne}, {Joshi}, {McLaughlin}, {Kramer}, {Sarkissian},
  {Camilo}, {Kalogera}, {Kim}, \& {Lorimer}}]{2003Natur.426..531B}
{Burgay} M. {et~al.}, 2003, \nat, 426, 531

\bibitem[{{Cameron} {et~al}\mbox{.}(2018){Cameron}, {Champion}, {Kramer},
  {Bailes}, {Barr}, {Bassa}, {Bhandari}, {Bhat}, {Burgay}, {Burke-Spolaor},
  {Eatough}, {Flynn}, {Freire}, {Jameson}, {Johnston}, {Karuppusamy}, {Keith},
  {Levin}, {Lorimer}, {Lyne}, {McLaughlin}, {Ng}, {Petroff}, {Possenti},
  {Ridolfi}, {Stappers}, {van Straten}, {Tauris}, {Tiburzi}, \&
  {Wex}}]{2018MNRAS.475L..57C}
{Cameron} A.~D. {et~al.}, 2018, \mnras, 475, L57

\bibitem[{{Ciufolini} {et~al}\mbox{.}(2013){Ciufolini}, {Paolozzi}, {Koenig},
  {Pavlis}, {Ries}, {Matzner}, {Gurzadyan}, {Penrose}, {Sindoni}, \&
  {Paris}}]{2013NuPhS.243..180C}
{Ciufolini} I. {et~al.}, 2013, Nuclear Physics B Proceedings Supplements, 243,
  180

\bibitem[{{Damour} \& {Ruffini}(1974)}]{1974CRASM.279..971D}
{Damour} T., {Ruffini} R., 1974, C.R. Acad. Sc. Paris, S\'{e}rie A, 279, 971

\bibitem[{{Damour} \& {Sch{\"a}fer}(1988)}]{1988NCimB.101..127D}
{Damour} T., {Sch{\"a}fer} G., 1988, Nuovo Cimento B, 101, 127

\bibitem[{{Damour} \& {Taylor}(1992)}]{1992PhRvD..45.1840D}
{Damour} T., {Taylor} J.~H., 1992, Phys. Rev. D, 45, 1840

\bibitem[{{Einstein}(1915)}]{Ein15}
{Einstein} A., 1915, Sitzungsber. K\"{o}n. Preu{\ss}. Akad. Wiss. zu Berlin,
  831–839

\bibitem[{{Everitt}(1974)}]{Varenna74}
{Everitt} C.~W.~F., 1974, in Proceedings of the International School of Physics
  \virg{Enrico Fermi}. Course LVI. Experimental Gravitation, {Bertotti} B.,
  ed., Academic Press, New York and London, pp. 331--360

\bibitem[{{Everitt} {et~al}\mbox{.}(2001){Everitt}, {Buchman}, {Debra},
  {Keiser}, {Lockhart}, {Muhlfelder}, {Parkinson}, \&
  {Turneaure}}]{2001LNP...562...52E}
{Everitt} C.~W.~F., {Buchman} S., {Debra} D.~B., {Keiser} G.~M., {Lockhart}
  J.~M., {Muhlfelder} B., {Parkinson} B.~W., {Turneaure} J.~P., 2001, in
  Lecture Notes in Physics, Berlin Springer Verlag, Vol. 562, Gyros, Clocks,
  Interferometers ...: Testing Relativistic Gravity in Space, {L{\"a}mmerzahl}
  C., {Everitt} C.~W.~F., {Hehl} F.~W., eds., p.~52

\bibitem[{{Everitt} {et~al}\mbox{.}(2011){Everitt}, {Debra}, {Parkinson},
  {Turneaure}, {Conklin}, {Heifetz}, {Keiser}, {Silbergleit}, {Holmes},
  {Kolodziejczak}, {Al-Meshari}, {Mester}, {Muhlfelder}, {Solomonik}, {Stahl},
  {Worden}, {Bencze}, {Buchman}, {Clarke}, {Al-Jadaan}, {Al-Jibreen}, {Li},
  {Lipa}, {Lockhart}, {Al-Suwaidan}, {Taber}, \& {Wang}}]{2011PhRvL.106v1101E}
{Everitt} C.~W.~F. {et~al.}, 2011, Phys. Rev. Lett., 106, 221101

\bibitem[{{Everitt} {et~al}\mbox{.}(2015){Everitt}, {Muhlfelder}, {DeBra},
  {Parkinson}, {Turneaure}, {Silbergleit}, {Acworth}, {Adams}, {Adler},
  {Bencze}, {Berberian}, {Bernier}, {Bower}, {Brumley}, {Buchman}, {Burns},
  {Clarke}, {Conklin}, {Eglington}, {Green}, {Gutt}, {Gwo}, {Hanuschak}, {He},
  {Heifetz}, {Hipkins}, {Holmes}, {Kahn}, {Keiser}, {Kozaczuk}, {Langenstein},
  {Li}, {Lipa}, {Lockhart}, {Luo}, {Mandel}, {Marcelja}, {Mester}, {Ndili},
  {Ohshima}, {Overduin}, {Salomon}, {Santiago}, {Shestople}, {Solomonik},
  {Stahl}, {Taber}, {Van Patten}, {Wang}, {Wade}, {Worden}, {Bartel}, {Herman},
  {Lebach}, {Ratner}, {Ransom}, {Shapiro}, {Small}, {Stroozas}, {Geveden},
  {Goebel}, {Horack}, {Kolodziejczak}, {Lyons}, {Olivier}, {Peters}, {Smith},
  {Till}, {Wooten}, {Reeve}, {Anderson}, {Bennett}, {Burns}, {Dougherty},
  {Dulgov}, {Frank}, {Huff}, {Katz}, {Kirschenbaum}, {Mason}, {Murray},
  {Parmley}, {Ratner}, {Reynolds}, {Rittmuller}, {Schweiger}, {Shehata},
  {Triebes}, {VandenBeukel}, {Vassar}, {Al-Saud}, {Al-Jadaan}, {Al-Jibreen},
  {Al-Meshari}, \& {Al-Suwaidan}}]{2015CQGra..32v4001E}
{Everitt} C.~W.~F. {et~al.}, 2015, Class. Quantum Gravity, 32, 224001

\bibitem[{{Ferdman} {et~al}\mbox{.}(2013){Ferdman}, {Stairs}, {Kramer},
  {Breton}, {McLaughlin}, {Freire}, {Possenti}, {Stappers}, {Kaspi},
  {Manchester}, \& {Lyne}}]{2013ApJ...767...85F}
{Ferdman} R.~D. {et~al.}, 2013, \apj, 767, 85

\bibitem[{{Hu} {et~al}\mbox{.}(2020){Hu}, {Kramer}, {Wex}, {Champion}, \&
  {Kehl}}]{2020MNRAS.497.3118H}
{Hu} H., {Kramer} M., {Wex} N., {Champion} D.~J., {Kehl} M.~S., 2020, \mnras,
  497, 3118

\bibitem[{{Iorio}(2009)}]{2009NewA...14...40I}
{Iorio} L., 2009, \na, 14, 40

\bibitem[{{Iorio}(2017)}]{2017EPJC...77..439I}
{Iorio} L., 2017, Eur. Phys. J. C, 77, 439

\bibitem[{{Iorio} {et~al}\mbox{.}(2011){Iorio}, {Lichtenegger}, {Ruggiero}, \&
  {Corda}}]{2011Ap&SS.331..351I}
{Iorio} L., {Lichtenegger} H. I.~M., {Ruggiero} M.~L., {Corda} C., 2011, \apss,
  331, 351

\bibitem[{{Kalitzin}(1959)}]{Kali59}
{Kalitzin} N.~S., 1959, N. Cimento, 11, 178

\bibitem[{{Kehl} {et~al}\mbox{.}(2017){Kehl}, {Wex}, {Kramer}, \&
  {Liu}}]{Kehletal017}
{Kehl} M.~S., {Wex} N., {Kramer} M., {Liu} K., 2017, in The Fourteenth Marcel
  Grossmann Meeting. Proceedings of the MG14 Meeting on General Relativity,
  {Bianchi} M., {Jantzen} R., {Ruffini} R., eds., World Scientific, Singapore,
  pp. 1860--1865

\bibitem[{{Kramer} {et~al}\mbox{.}(2006){Kramer}, {Stairs}, {Manchester},
  {McLaughlin}, {Lyne}, {Ferdman}, {Burgay}, {Lorimer}, {Possenti}, {D'Amico},
  {Sarkissian}, {Hobbs}, {Reynolds}, {Freire}, \&
  {Camilo}}]{2006Sci...314...97K}
{Kramer} M. {et~al.}, 2006, Science, 314, 97

\bibitem[{{Kramer} {et~al}\mbox{.}(2021){Kramer}, {Stairs}, {Venkatraman
  Krishnan}, {Freire}, {Abbate}, {Bailes}, {Burgay}, {Buchner}, {Champion},
  {Cognard}, {Gautam}, {Geyer}, {Guillemot}, {Hu}, {Janssen}, {Lower},
  {Parthasarathy}, {Possenti}, {Ransom}, {Reardon}, {Ridolfi}, {Serylak},
  {Shannon}, {Spiewak}, {Theureau}, {van Straten}, {Wex}, {Oswald}, {Posselt},
  {Sobey}, {Barr}, {Camilo}, {Hugo}, {Jameson}, {Johnston}, {Karastergiou},
  {Keith}, \& {Os{\l}owski}}]{2021MNRAS.504.2094K}
{Kramer} M. {et~al.}, 2021, \mnras, 504, 2094

\bibitem[{{Laarakkers} \& {Poisson}(1999)}]{1999ApJ...512..282L}
{Laarakkers} W.~G., {Poisson} E., 1999, \apj, 512, 282

\bibitem[{{Lattimer} \& {Schutz}(2005)}]{2005ApJ...629..979L}
{Lattimer} J.~M., {Schutz} B.~F., 2005, \apj, 629, 979

\bibitem[{{Lense} \& {Thirring}(1918)}]{LT18}
{Lense} J., {Thirring} H., 1918, PhyZ, 19, 156

\bibitem[{{Lucchesi} {et~al}\mbox{.}(2020){Lucchesi}, {Visco}, {Peron},
  {Bassan}, {Pucacco}, {Pardini}, {Anselmo}, \&
  {Magnafico}}]{2020Univ....6..139L}
{Lucchesi} D., {Visco} M., {Peron} R., {Bassan} M., {Pucacco} G., {Pardini} C.,
  {Anselmo} L., {Magnafico} C., 2020, Universe, 6, 139

\bibitem[{{Lucchesi} {et~al}\mbox{.}(2019){Lucchesi}, {Anselmo}, {Bassan},
  {Magnafico}, {Pardini}, {Peron}, {Pucacco}, \& {Visco}}]{2019Univ....5..141L}
{Lucchesi} D.~M., {Anselmo} L., {Bassan} M., {Magnafico} C., {Pardini} C.,
  {Peron} R., {Pucacco} G., {Visco} M., 2019, Universe, 5, 141

\bibitem[{{Lyne}(1984)}]{1984Natur.310..300L}
{Lyne} A.~G., 1984, \nat, 310, 300

\bibitem[{{Lyne} {et~al}\mbox{.}(2004){Lyne}, {Burgay}, {Kramer}, {Possenti},
  {Manchester}, {Camilo}, {McLaughlin}, {Lorimer}, {D'Amico}, {Joshi},
  {Reynolds}, \& {Freire}}]{2004Sci...303.1153L}
{Lyne} A.~G. {et~al.}, 2004, Science, 303, 1153

\bibitem[{{Mashhoon}(2001)}]{2001rfg..conf..121M}
{Mashhoon} B., 2001, in Reference Frames and Gravitomagnetism,
  {Pascual-S{\'a}nchez} J.~F., {Flor{\'\i}a} L., {San Miguel} A., {Vicente} F.,
  eds., World Scientific, Singapore, pp. 121--132

\bibitem[{{Mashhoon}(2003)}]{Mash03}
{Mashhoon} B., 2003, in The Measurement of Gravitomagnetism: A Challenging
  Enterprise, {Iorio} L., ed., Nova Science Publishers, Hauppauge (NY), pp.
  29--39

\bibitem[{{Mashhoon}, {Hehl} \& {Theiss}(1984){Mashhoon}, {Hehl}, \&
  {Theiss}}]{1984GReGr..16..727M}
{Mashhoon} B., {Hehl} F.~W., {Theiss} D.~S., 1984, General Relativity and
  Gravitation, 16, 727

\bibitem[{{Miao} {et~al}\mbox{.}(2021){Miao}, {Xu}, {Shao}, {Liu}, \&
  {Ma}}]{2021arXiv210705812M}
{Miao} X., {Xu} H., {Shao} L., {Liu} C., {Ma} B.-Q., 2021, arXiv e-prints,
  arXiv:2107.05812

\bibitem[{{Michalska}(1960)}]{Micha60}
{Michalska} R., 1960, Bull. Acad. Polon. Sci., Ser. Math. Astr. Phys., 8, 247

\bibitem[{{Pfister}(2007)}]{2007GReGr..39.1735P}
{Pfister} H., 2007, General Relativity and Gravitation, 39, 1735

\bibitem[{{Pfister}(2008)}]{2008mgm..conf.2456P}
{Pfister} H., 2008, in The Eleventh Marcel Grossmann Meeting On Recent
  Developments in Theoretical and Experimental General Relativity, Gravitation
  and Relativistic Field Theories, {Kleinert} H., {Jantzen} R.~T., {Ruffini}
  R., eds., World Scientific, Singapore, pp. 2456--2458

\bibitem[{{Pfister}(2014)}]{Pfister2014}
{Pfister} H., 2014, in Springer Proceedings in Physics, Vol. 157, Relativity
  and Gravitation, {Bi\v{c}\'{a}k} J., {Ledvinka} T., eds., Springer Verlag,
  Berlin/Heidelberg, pp. 191--197

\bibitem[{{Pugh}(1959)}]{Pugh59}
{Pugh} G., 1959, {Proposal for a Satellite Test of the Coriolis Prediction of
  General Relativity}. Research Memorandum~11, Weapons Systems Evaluation
  Group, The Pentagon, Washington D.C.

\bibitem[{{Renzetti}(2013{\natexlab{a}})}]{2013NewA...23...63R}
{Renzetti} G., 2013{\natexlab{a}}, \na, 23, 63

\bibitem[{{Renzetti}(2013{\natexlab{b}})}]{2013CEJPh..11..531R}
{Renzetti} G., 2013{\natexlab{b}}, Central European Journal of Physics, 11, 531

\bibitem[{{Renzetti}(2014)}]{2014NewA...29...25R}
{Renzetti} G., 2014, \na, 29, 25

\bibitem[{{Rickett} {et~al}\mbox{.}(2014){Rickett}, {Coles}, {Nava},
  {McLaughlin}, {Ransom}, {Camilo}, {Ferdman}, {Freire}, {Kramer}, {Lyne}, \&
  {Stairs}}]{2014ApJ...787..161R}
{Rickett} B.~J. {et~al.}, 2014, \apj, 787, 161

\bibitem[{{Rindler}(2001)}]{2001rsgc.book.....R}
{Rindler} W., 2001, {Relativity: special, general, and cosmological}. Oxford
  University Press, Oxford

\bibitem[{{Robertson}(1938)}]{Rob38}
{Robertson} H., 1938, Annals of Mathematics, 39, 101

\bibitem[{{Sch{\"a}fer} \& {Wex}(1993{\natexlab{a}})}]{1993PhLA..174..196S}
{Sch{\"a}fer} G., {Wex} N., 1993{\natexlab{a}}, Physics Letters A, 174, 196

\bibitem[{{Sch{\"a}fer} \& {Wex}(1993{\natexlab{b}})}]{1993PhLA..177..461S}
{Sch{\"a}fer} G., {Wex} N., 1993{\natexlab{b}}, Physics Letters A, 177, 461

\bibitem[{{Schiff}(1960)}]{Schiff60}
{Schiff} L., 1960, Phys. Rev. Lett., 4, 215

\bibitem[{{Silva} {et~al}\mbox{.}(2021){Silva}, {Holgado},
  {C{\'a}rdenas-Avenda{\~n}o}, \& {Yunes}}]{2021PhRvL.126r1101S}
{Silva} H.~O., {Holgado} A.~M., {C{\'a}rdenas-Avenda{\~n}o} A., {Yunes} N.,
  2021, \prl, 126, 181101

\bibitem[{{Smarr} \& {Blandford}(1976)}]{1976ApJ...207..574S}
{Smarr} L.~L., {Blandford} R., 1976, \apj, 207, 574

\bibitem[{{Soffel}(1989)}]{Sof89}
{Soffel} M.~H., 1989, Relativity in Astrometry, Celestial Mechanics and
  Geodesy. Springer, Heidelberg

\bibitem[{{Thorne}(1986)}]{1986hmac.book..103T}
{Thorne} K.~S., 1986, in Highlights of Modern Astrophysics: Concepts and
  Controversies, {Shapiro} S.~L., {Teukolsky} S.~A., {Salpeter} E.~E., eds.,
  Wiley, NY, p. 103

\bibitem[{{Thorne}(1988)}]{1988nznf.conf..573T}
{Thorne} K.~S., 1988, in Near Zero: New Frontiers of Physics, {Fairbank} J.~D.,
  {Deaver} Jr. B.~S., {Everitt} C.~W.~F., {Michelson} P.~F., eds., pp. 573--586

\bibitem[{{Thorne}, {MacDonald} \& {Price}(1986){Thorne}, {MacDonald}, \&
  {Price}}]{Thorne86}
{Thorne} K.~S., {MacDonald} D.~A., {Price} R.~H., eds., 1986, {Black Holes: The
  Membrane Paradigm}. Yale University Press, Yale

\bibitem[{{Wex}(1995)}]{1995CQGra..12..983W}
{Wex} N., 1995, Classical and Quantum Gravity, 12, 983

\bibitem[{{Wex}(1998)}]{1998MNRAS.298...67W}
{Wex} N., 1998, \mnras, 298, 67

\end{thebibliography}

\end{document}